\documentclass{aa}

\usepackage{graphicx}
\usepackage{txfonts}
\usepackage[svgnames]{xcolor}
\usepackage{natbib}
\bibpunct{(}{)}{;}{a}{}{,} 

\begin{document}

\title{Effects of the variation of fundamental constants on Pop III stellar evolution}
\titlerunning{Effect of the variation of fundamental constants on Pop III star evolution}
   \author{Sylvia Ekstr\"om\inst{1},
          Alain Coc\inst{2},       
          Pierre Descouvemont\inst{3},
          Georges Meynet\inst{1}
          Keith A. Olive\inst{4},
          Jean-Philippe Uzan\inst{5,6}
          \and
          Elisabeth Vangioni\inst{5}}
   \authorrunning{Ekstr\"om et al.}
   \offprints{Sylvia.Ekstrom@unige.ch}
   \institute{
   Geneva Observatory, University of Geneva, Maillettes 51, 1290 Sauverny, Switzerland
    \and 
    Centre de Spectrom\'etrie Nucl\'eaire et de Spectrom\'etrie de Masse (CSNSM), UMR 8609, CNRS/IN2P3 and Universit\'e 
          Paris Sud 11, B\^atiment 104, 91405 Orsay Campus, France
    \and 
    Physique Nucl\'eaire Th\'eorique et Physique Math\'ematique, C.P. 229, 
    Universit\'e Libre de Bruxelles (ULB), B-1050 Brussels, Belgium
    \and 
    William I. Fine Theoretical Physics Institute, University of Minnesota, 
     Minneapolis, Minnesota 55455, USA
    \and 
    Institut d'Astrophysique de Paris, UMR-7095 du CNRS, Universit\'e Pierre et Marie Curie, 
    98 bis bd Arago, 75014 Paris, France
    \and
    Department of Mathematics and Applied Mathematics, University of Cape Town,
    Rondebosch 7701, Cape Town, South Africa}
   
   \date{}

  \abstract
   {}
   {A variation of the fundamental constants is expected to affect the
 thermonuclear rates important for stellar nucleosynthesis. In particular, because of the very small resonant energies of $^8$Be  and $^{12}$C,  the triple $\alpha$ process is extremely sensitive to any such variations. 
}
  {Using a microscopic model for these nuclei, we derive the sensitivity of the Hoyle state
to the nucleon-nucleon potential  allowing for a change in the magnitude of the nuclear interaction.
We follow the evolution of 15 and 60 $M_{\sun}$ zero metallicity stellar models, up to the end of core helium burning. These stars are assumed to be representative of the first, Population III stars.
}
  {We derive limits on the variation of the magnitude of the nuclear interaction and model dependent limits on the variation of 
the fine structure constant based on  
the calculated oxygen and carbon abundances resulting from helium burning. 
The requirement that some $^{12}$C and $^{16}$O be present are the end of the helium burning phase
allows for permille limits on the change of the nuclear interaction and limits of order $10^{-5}$ on the fine structure constant relevant at a cosmological redshift of $z \sim 15-20$. 
}
  {}

   \keywords{Stars: evolution, population III, nucleosynthesis -- Fundamental constants: variation -- 
   General Relativity: tests -- Equivalence principle}

   \maketitle
\section{Introduction}

The equivalence principle is a cornerstone of metric theories of gravitation and in particular of general relativity \citep{willbook}. This principle, including the universality of free fall, the local position and Lorentz invariances, postulates that the local laws of physics, and in particular the values of the dimensionless constants such as the fine structure constant $\alpha_{em}\equiv e^2/4\pi\varepsilon_0\hbar c$, must remain fixed, and thus be the same at  any time and in any place. It follows that by testing the constancy of fundamental constants one actually performs a test of General Relativity, that can be extended on astrophysical and cosmological scales~\citep[for a review, see][]{uzanrevue1,Uzan2009a}

We define a fundamental constant as any free parameter of the fundamental theories at hand~\citep{weinberg83,duff,tri,barrowbook,jpbook}. These parameters are contingent quantities that can only be measured and are assumed constant since (i) in the theoretical framework in which they appear, there is  no equation of motion for them and they cannot be deduced from other constants and (ii) if the theories in which they appear have been validated experimentally, it means that, at the precision of the experiments, these parameters have indeed been checked to be constant. Hence, by testing for their constancy we extend our knowledge of the domain of validity of the theories in which they appear. In that respect, astrophysics and cosmology allow one to probe larger time-scales, typically of the order of the age of the universe.

One can, however, question the constancy of these dimensionless numbers and the physics which determines their value. This sends us back to the phenomenological argument by~\citet{dirac37}, known as the `Large Number Hypothesis', according to which the dimensionless ratio $Gm_{\rm e}m_{\rm p}/\hbar c$, or  simply $G$ in atomic units, should decrease as the inverse of the age of the universe, followed by~\citet{jordan37} who formulated a field theory in which both the fine structure constant and the gravitational constant were replaced by dynamical fields. It was soon pointed out by~\citet{fierz56} that astronomical observations can set strong constraints on the variations of these constants. This paved the way to two complementary directions of research on the fundamental constants.

On the one hand, from a theoretical perspective, many theories involving ``varying constants'' have been designed. This is in particular the case of theories involving extra-dimensions, such as the Kaluza-Klein mechanism~\citep{kaluza21,Klein26} and string theory, in which all the constants (including gauge, Yukawa and gravitational couplings) are dynamical quantities \citep{Wu:1986ac,Wetterich:1987fk, taylor, witten}, or in theories such as scalar-tensor theories of gravity~\citep{Jordan49,bd61,gefdam} and in many models of quintessence~\citep{uzan99,dam02, rundil,wetterich,Lee:2004vm, riaz}, aiming at explaining the acceleration of the universe by the dynamics of a scalar field. It is impingent on these models to explain why the constants are so constant today and provide a mechanism for fixing their value~\citep{dn,dp}. In this respect, testing for the constancy of the fundamental constants is one of the few windows on these theories.

On the other hand, from an experimental and observational perspective, the variation of various constants have been severely constrained. This is the case of the fine structure constant for which the constraint $\dot\alpha_{em}/\alpha_{em}=(-1.6\pm2.3)\times10^{-17}\,{\rm yr}^{-1}$ at $z=0$ has been obtained from the comparison of aluminium and mercury single-ion optical clocks~\citep{rosenband}. Over a longer timescale, it was demonstrated that $\alpha_{em}$ cannot have varied by more than $10^{-7}$ over the last 2 Gyr  from the Oklo phenomenon~\citep{oklo0,oklo1,Fujii:1998kn,oklo2,oklo3} and over the last 4.5 Gyr from meteorite dating~\citep{meteo1,dyson,meteo2,meteo3}. At higher redshift, $0.4 < z < 3.5$, there are conflicting reports of a an observed variation of $\alpha_{em}$ from quasar absorption systems.  Using the many-multiplet method, \citet{webb01} and \cite{murph03,murphy07} claim a statistically significant variation $\Delta\alpha_{em}/\alpha_{em} = (-0.54 \pm 0.12)\times 10^{-5}$, indicating a smaller value of $\alpha_{em}$ in the past. More recent observations taken at VLT/UVES using the many multiplet method have not been able to duplicate the previous result \citep{chand,sri04,quast04,srianand2007}. The use of Fe lines in \citet{quast04} on a single absorber found $\Delta \alpha_{em} / \alpha_{em} = (-0.05 \pm 0.17) \times 10^{-5}$.  However, since the previous result relied on a statistical average of over 100 absorbers, it is not clear that these two results are in contradiction.  In \citet{chand}, the use of Mg and Fe lines in a set of 23 systems yielded the result $\Delta \alpha_{em} / \alpha_{em} = (0.01 \pm 0.15) \times 10^{-5}$ and therefore represents a more significant disagreement and can be used to set very stringent limits on the possible variation in $\alpha_{em}$. A purely astrophysical explanation for these results is also possible \citep{amo,amo2}. At larger redshifts, constraints at the percent level have been obtained from the observation of the temperature anisotropies of cosmic microwave background at ($z\sim10^{3}$) \citep[\textsl{e.g. }][]{cmb4,cmb3,cmb1,cmb2} and from big bang  nucleosynthesis (BBN)  ($z\sim10^{10}$) \citep[\textsl{e.g. }][]{kpw,campolive95,Bergstrom,Nollett,Ichikawa02,bbn2,bbn3,Ichikawa04,bbn4,coc07,bbn1}. We refer to \citet{uzanrevue1,uzanrevue2,uzanrevue3} for recent reviews on this topic. For the time being, there is no constraint on $\alpha_{em}$ for redshifts ranging from 4 to $10^3$ although it has been proposed that 21~cm observations may allow one  to fill in the range $30<z<100$~\citep{21cm}.

This article focuses on the effect of the possible variation of the fundamental constants on the stellar evolution of early stars, hence possibly providing constraints in a domain of redshifts where no such constraint is available. A similar issue was actually considered by~\cite{gamow67} (see also the recent work by~\citet{adams} who showed that the evolution of the Sun was able to exclude the Dirac model of a varying gravitational constant. In this case, non-gravitational physics is kept unchanged and the evolution of the star is affected only by the modification of gravity. Changing the non-gravitational sector has more drastic implications on stellar physics since the nuclear physics and thus the cross-sections and reaction rates of all the processes should be modified.

\citet{rozen88} argued that the synthesis of complex elements in stars (mainly the possibility of the $3\alpha$-reaction as the origin of the production of ${}^{12}{\rm C}$) sets constraints on the values of the fine structure and strong coupling constants. There have been several studies on the sensitivity of carbon production to the underlying nuclear rates \citep{livio,Fairbairn,csoto2000,ober2000,ober2002,schlattl2004,tur}. The production of ${}^{12}{\rm C}$ in stars requires a triple tuning: (i) the decay lifetime of ${}^8{\rm Be}$, of order $10^{-16}$~s, is four orders of magnitude longer than the time for two $\alpha$ particles to scatter, (ii) an excited state of the carbon lies just above the energy of ${}^8{\rm Be}+\alpha$ and finally (iii) the energy level of ${}^{16}{\rm O}$ at 7.1197~MeV is non resonant and below the energy of ${}^{12}{\rm C}+\alpha$, at 7.1616~MeV, which ensures that most of the carbon synthesized is not destroyed by the capture of an $\alpha$-particle. The existence of this excited state of $^{12}$C was actually predicted by~\citet{hoyle} and then observed at the predicted energy by~\citet{dunbar} as well as its decay~\citep{cook}. The variation of any constant which would modify the energy of this resonance, known as the Hoyle level, would dramatically affect the production of carbon. 

Qualitatively, and perhaps counter-intuitively, if the energy level of the Hoyle level were increased, $^{12}$C would probably be rapidly processed to $^{16}$O since the star would, in fact, need to be hotter for the triple-$\alpha$ reaction to be triggered. On the other hand, if it is decreased very little oxygen will be produced. From the general expression of the reaction rate (see below for details, definitions of all the quantities entering this expression, and a more accurate computation)
$$
 \lambda_{3\alpha}=3^{3/2}N_\alpha^3 \left(\frac{2\pi\hbar^3}{M_\alpha k_{\rm B}T}\right)^{3}\frac{\Gamma}{\hbar} \exp\left[-\frac{Q_{\alpha\alpha\alpha}}{k_{\rm B}T}\right],
$$
where $Q_{\alpha\alpha\alpha}~\sim380$~keV is the energy of the resonance, one deduces that the sensitivity of the reaction rate to a variation of $Q_{\alpha\alpha\alpha}$ is
$$
 s=\frac{{\rm d}\ln\lambda_{3\alpha}}{{\rm d}\ln Q_{\alpha\alpha\alpha}} =-\frac{Q_{\alpha\alpha\alpha}}{k_{\rm B}T}\sim  \left(\frac{-4.4}{T_9}\right).
$$
where $T_9 = T/10^9$K. This effect was investigated by~\citet{csoto2000} and \cite{ober2000,ober2002} who related the variation of $Q_{\alpha\alpha\alpha}$ to a variation of the strength of the nucleon-nucleon (N-N) interaction. Focusing on the C/O ratio in red giant stars (1.3, 5 and 20 M$_{\sun}$ with solar metallicity) up to thermally pulsing asymptotic giant branch stars (TP-AGB)~\citep{ober2000,ober2002} and in low, intermediate and high mass stars (1.3, 5, 15 and 25 M$_{\sun}$ with solar metallicity)~\citep{schlattl2004}, it was estimated that outside a window of 0.5\% and 4\% for the values of the strong and electromagnetic forces respectively, the stellar production of carbon or oxygen will be reduced by a factor 30 to 1000 (see also \citet{pochet}). 

Indeed, modifying the energy of the resonance alone is not realistic since all cross-sections, reaction rates and binding energies etc. should be affected by the variation of the constants. One could indeed have started by assuming independent variations of all these quantities but it is more realistic (and hence more model-dependent) to try to deduce their variation from a microscopic model.  Our analysis can then be outlined in three main steps:
\begin{enumerate}
 \item Relating the nuclear parameters to fundamental constants such as the Yukawa and gauge couplings, and the Higgs vacuum expectation value. This is a difficult step because of the intricate structure of QCD and its role in low energy nuclear reactions, as in the case of BBN. The nuclear parameters include the set of relevant energy levels (including the ground states), binding energies of each nucleus and the partial width of each nuclear reaction. This involves a nuclear physics model of the relevant nuclei (mainly $^4$He, $^8$Be, $^{12}$C, and $^{16}$O for our study).
 \item Relating the reaction rates to the nuclear parameters, which implies an integration over energy of the cross-sections.
 \item Deducing the change in the stellar evolution (lifetime of the star, abundance of the nuclei, Hertzprung-Russel (H-R) diagram, etc.). This involves a stellar model.
\end{enumerate}

Let us summarize the main hypothesis of our work for each of these steps.

The {\bf first step} is probably the most difficult. We shall adopt a phenomenological description of the different nuclei based on a cluster model in which the wave functions of the $^8$Be and $^{12}$C nuclei are approximated by a cluster of respectively two and three $\alpha$ wave functions. When solving the associated Schr\"odinger equation, we will modify the strength of the electromagnetic and nuclear N-N interaction potentials respectively by a factor  $(1+\delta_\alpha)$ and $(1+\delta_{NN})$ where $\delta_\alpha$ and $\delta_{NN}$ are two small dimensionless parameters that encode the variation of the fine structure constant and other fundamental couplings. At this stage, the relation between $\delta_{NN}$ and the gauge and Yukawa couplings is not known. This will allow us to obtain the energy levels, including the binding energy, of $^2$H, $^4$He, $^8$Be, $^{12}$C and the first  $J^\pi$ = 0$^+$ $^{12}$C excited energy level. Note that all of the relevant nuclear states are assumed to be interacting alpha clusters. In a first approximation, the variation of the $\alpha$ particle mass cancels out. The partial widths (and lifetimes) of these states are scaled from their experimental laboratory values, according to their energy dependence. $\delta_{NN}$ is used as a free parameter. The dependence of the deuterium binding energy on $\delta_{NN}$ then offers us the possibility of relating this parameter to the gauge and Yukawa couplings if one matches this prediction to a potential model via the $\sigma$ and $\omega$ meson masses~\citep{sig1,sig2,coc07, dd} or the pion mass, as suggested by \citet{pi1,pi2,pi3}.

The {\bf second step} requires an integration over energy to deduce the reaction rates as functions of the temperature and of the new parameters $\delta_\alpha$ and $\delta_{NN}$.

The {\bf third step} involves stellar models and in particular some choices about the masses and initial metallicity of the stars. While theoretically uncertain, it is usually thought that the first stars were massive; however, their mass range is presently unknown, (for a review, see \cite{ bromm09}). In a hierarchical scenario of structure formation, they were formed a few $\times~10^8$ years after the big bang, that is at a redshift of $z\sim10-15$ with zero metallicity (so that we can use the BBN abundances as initial conditions, see Sect.~\ref{sec_stellar}). We thus focus on Population~III stars (Pop III) with typical masses, 15 and 60 $M_{\sun}$, assuming no rotation. Our computation is stopped at the end of core helium burning.

The {\bf final step} would be to use these predictions to set constraints on the fundamental constants, using stellar constraints such the C/O ratio which is in fact observable in very metal poor stars. While this article can be seen as a theoretical investigation that describes the expected effect of a variation of the fundamental constants, it also sheds some interesting light on stellar physics and its sensitivity to fundamental physics.

The article is logically organized as follows. Section 2 recalls the basis of the $3\alpha$-reaction, Section 3 describes the nuclear physics modeling (first step), Section 4 is devoted to stellar implications and Section 5 to the discussion. Technical details are gathered in the appendices.

\section{Stellar carbon production}

The $3\alpha$  process is one of the most delicate of all reactions in nuclear astrophysics. It is also one of the most influential since it bypasses the deep gap between BBN and stellar nucleosynthesis. More specifically, BBN stops at mass 7 ($^7$Li) due to the lethal instability of parent nuclei with strongly bound$^4$He offsprings, namely nuclei with masses  5 and 8. The  $3\alpha$ reaction allows nuclear complexity to proceed up to uranium through core collapse supernova explosions. Note that the gap between BBN and stellar nucleosynthesis is filled by non-thermal processes (spallative processes induced by galactic cosmic rays) producing $^{6,7}$Li, $^9$Be and $^{10,11}$B. Once these nuclear obstacles are overcome, the physical conditions within stars allow nuclear production  of elements from carbon and beyond. As such, the $3\alpha$ reaction is the first step of helium burning which is followed in massive stars by C, Ne, O, and Si burning and then explosive nucleosynthesis. The mass of the C-O core and the C/O ratio at the end of helium burning is important for determining i) the subsequent phases of stellar evolution and nucleosynthesis since it fixes the mass of the iron core and ii) the final fate of stars (black holes, neutron stars, or white dwarfs). In particular for the first stars, the $3\alpha$ reaction is of great importance since no metals have yet been formed and the CNO cycle cannot proceed as usual. Unfortunately, the $3\alpha$ reaction is a two step sequential process and the $^8$Be($\alpha$,$\gamma$)$^{12}$C cross section has not been measured directly in the laboratory. Indeed, the $^8$Be lifetime (about 10$^{-16}$ sec) is so short that such a measurement is not currently feasible.
          
Consequently, the C and O abundances at the end of helium burning is very sensitive to small variations of the $3\alpha$ reaction rate. In this context, any anomalous abundance of C and O in very metal poor stars could potentially be taken as an indication of the variation of the nucleon - nucleon interaction and therefore of either or both of the electromagnetic and strong coupling constants. 
    
In our analysis, we focus on the C/O ratio. It is of interest, therefore, to comment on the destruction of carbon (production of oxygen) as well as the destruction of oxygen. If the reaction following the $3\alpha$ process, namely $^{12}$C($\alpha$, $\gamma$)$^{16}$O, is sufficiently fast, then most $\alpha$ particles would be converted to $^{16}$O or heavier nuclei with little $^{12}$C left at the end of helium burning. However, the fact that in general the C/O ratio in the Universe is about 0.4 suggests that the $^{12}$C($\alpha$, $\gamma$)$^{16}$O  reaction is sufficiently slow that some $^{12}$C remains after helium exhaustion. The presence of comparable quantities of C and O implies also that the subsequent $^{16}$O( $\alpha$, $\gamma$)$^{20}$Ne reaction is not too fast , otherwise O would be converted to Ne or heavier nuclei and little O would survive during helium burning. We would like to stress the importance of the nuclear  balance between C and O. The observation of C/O  in very metal poor stars may hold the key to any variation in the chain of processes described above. 

\begin{figure}
\centering
 \resizebox{.8\hsize}{!}{\includegraphics{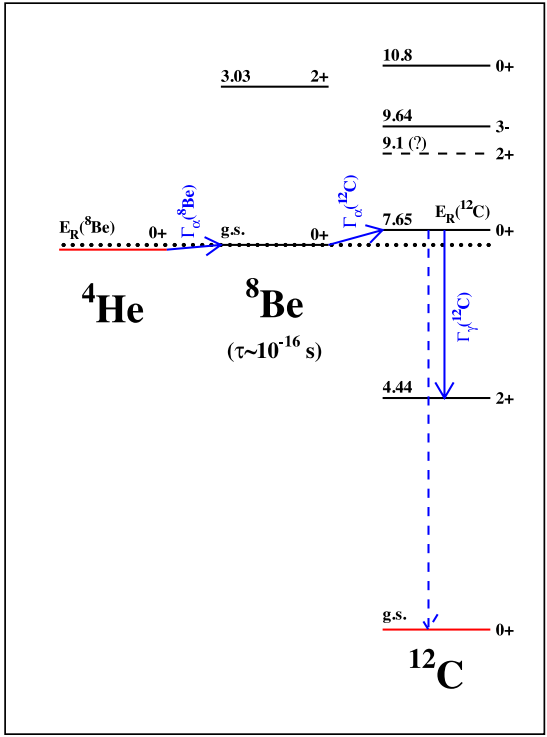}}
\caption{Level scheme showing the key levels in the triple $\alpha$ process.}
\label{flevels}
\end{figure}

In Fig.~\ref{flevels}, we show the low energy level schemes of the nuclei participating to the $^4$He($\alpha\alpha,\gamma$)$^{12}$C reaction: $^4$He, $^8$Be and $^{12}$C. The triple $\alpha$ process begins when two alpha particles fuse to produce a $^8$Be nucleus whose lifetime is only $\sim10^{-16}$~s but is sufficiently long so as to allow a second alpha capture into the second excited level of $^{12}$C, at 7.65~MeV above the ground state (of $^{12}$C). In the following, we shall refer to the successive $\alpha$ captures  as first and second steps, that is  $\alpha\alpha\leftrightarrow ^{8}$Be$+\gamma$ and $^{8}$Be$+\alpha\leftrightarrow ^{12}$C$^*\rightarrow ^{12}$C$+\gamma$. The excited state of $^{12}$C corresponds to an $\ell$ = 0 resonance, as postulated by~\cite{hoyle} in order to increase the cross section during the helium burning phase. This level decays to the first excited level of $^{12}$C at 4.44~MeV through an $E2$ (i.e. electric with $\ell$ = 2 multipolarity) radiative transition as the transition to the ground state ($0_1^+\to0_2^+$) is suppressed (pair emission only). At temperatures above $T_9  \approx 2$, which are not relevant for our analysis and therefore not treated, one should also consider other possible levels above the $\alpha$ threshold.  

We define the following energies:
\begin{itemize}
 \item $E_R$($^8$Be) as the energy of the $^8$Be ground state with respect to the $\alpha$+$\alpha$ threshold;
 \item $E_R$($^{12}$C) as the energy of the Hoyle level with respect to the $^8$Be+$\alpha$ threshold, i.e. $E_R$($^{12}$C) $\equiv$ $^{12}$C(0$_2$$^+$)+$Q_\alpha(^{12}$C) where  $^{12}$C(0$_2$$^+$) is the excitation energy and $Q_\alpha(^{12}$C) is the $\alpha$ particle separation energy;
 \item $Q_{\alpha\alpha\alpha}$ as the energy of the Hoyle level with respect to the $3 \alpha$ threshold so that
\begin{equation}\label{Q3alpha}
 Q_{\alpha\alpha\alpha}= E_R( ^8{\rm Be}) +E_R( ^{12}{\rm C}).
\end{equation}
 \item $\Gamma_\alpha$($^8$Be) as the partial width of the beryllium decay ($\alpha\alpha\leftrightarrow ^{8}$Be$+\gamma$);
 \item $\Gamma_{\gamma,\alpha}$($^{12}$C) as the partial widths of $^{8}$Be$+\alpha\leftrightarrow ^{12}$C$^*\rightarrow ^{12}$C$+\gamma$.
\end{itemize}
Their standard values are given in Table~\ref{t:widths}.

\begin{table}
\begin{center}
\caption{Nuclear data for the two steps of the $3\alpha$-reaction. See text for the definitions of the quantities \protect\citep{TUNL,Ajz90,audi03}.}\label{t:widths}
\begin{tabular}{c|c|c|c|c}
  \hline
  \hline
 nucleus&$J^{\pi}$&$E_R$ (keV)&$\Gamma_{\alpha}$ (eV)&
$\Gamma_{\gamma}$ (meV) \\
\hline
$^8$Be&$0^+$& 91.84$\pm$0.04&$5.57 \pm 0.25$   &$-$  \\
$^{12}$C&$0^+_2$&287.6$\pm$0.2& $8.3\pm1.0$ &$3.7\pm0.5$ \\
\hline
\end{tabular}
\end{center}
\end{table}

Assuming $i$) thermal equilibrium between the $^4$He\ and $^8$Be nuclei, so that their abundances are related by the Saha equation and $ii$) the sharp resonance approximation for the alpha capture on $^8$Be, the $^4$He($\alpha\alpha,\gamma$)$^{12}$C rate can be expressed \citep{Nom85,Il07} as: 
\begin{eqnarray}
N_{\rm A}^2 \langle \sigma v \rangle^{\alpha\alpha\alpha} = 3^{3/2}6N_{\rm A}^2
\left( \frac{2\pi}{M_\alpha k_{\rm B} T} \right)^3 
\hbar^5\omega\gamma
\exp\left({-{Q_{\alpha\alpha\alpha}}\over{k_{\rm B} T}}\right)
\label{eq:anal} 
\end{eqnarray}
with $\omega=1$ (spin factor), $\gamma=\Gamma_\gamma$($^{12}$C)$\Gamma_\alpha(^{12}$C)/$(\Gamma_\gamma(^{12}$C)+$\Gamma_\alpha(^{12}$C)) $\approx\Gamma_\gamma(^{12}$C) for present day values, and $M_\alpha$ is the mass of the $\alpha$ nucleus. 

During helium burning, the only other important reaction is  $^{12}$C$(\alpha,\gamma) ^{16}$O \citep{Il07} which transforms $^{12}$C into $^{16}$O. Its competition with the $3\alpha$ reaction governs the $^{12}$C/$^{16}$O abundance ratio at the end of the helium burning phase. Even though, the precise value of the $^{12}$C$(\alpha,\gamma) ^{16}$O $\it{S}$--factor\footnote{The astrophysical $\it{S}$--factor is just the cross section corrected for the effect of the penetrability of the Coulomb barrier and other trivial effects.} is still a matter of debate as it relies on an extrapolation of experimental data down to the astrophysical energy ($\approx$300~keV), its energy dependence is much weaker than fthat of the $3\alpha$ reaction. Indeed, as it is dominated by {\em broad} resonances, a shift of a few hundred keV in energy results in a  $\it{S}$--factor variation of much less than an order of magnitude. For this reason, we can safely neglect the effect of  the $^{12}$C$(\alpha,\gamma) ^{16}$O reaction rate variation when compared to the variation in the $3\alpha$ rate. Similar considerations apply to the rate for $^{16}$O$(\alpha,\gamma)^{20}$Ne. 

During hydrogen burning, the pace of the CNO cycle is given by the slowest reaction, $^{14}$N(p,$\gamma)^{15}$O. Its $\it{S}$--factor exhibits a well known resonance at 260~keV which is normally outside of the Gamow energy window ($\approx$100 keV) but a variation of the N--N potential could shift its position downward, resulting in a higher reaction rate and more efficient CNO H-- burning.

\section{Microscopic determination of the \boldmath $3\alpha$ rate \label{s3alpha}}
\subsection{Description of the cluster model}

In order to analyze the sensitivity of the triple $\alpha$ reaction to a variation of the strength of the electromagnetic and NN interactions, we use a microscopic model \citep[see][and references therein]{WT77,KD04}. In such an approach, the wave function of a  nucleus with atomic number $A$, spin $J$, and total parity $\pi$ is a solution of a Schr{\"o}dinger equation with a Hamiltonian given by
\begin{eqnarray}\label{eq1}
H=\sum_i^A T(\mathbf{r}_i)+\sum_{i>j=1}^A V(\mathbf{r}_{ij}).
\end{eqnarray}
$T(\mathbf{r}_i)$ is  the kinetic energy of nucleon $i$. The nucleon-nucleon interaction $V(\mathbf{r}_{ij})$ depends only on the set of relative distances $\mathbf{r}_{ij}=\mathbf{r}_i-\mathbf{r}_j$. It can be decomposed as
\begin{eqnarray}\label{eq2}
 V(\mathbf{r}_{ij})=V_C(\mathbf{r}_{ij})+V_N(\mathbf{r}_{ij}),
\end{eqnarray}
where the potential $V_C(\mathbf{r})$ arises from the electromagnetic interaction and $V_N(\mathbf{r})$ from the nuclear interaction. The expression for $V_N$ is detailed in Appendix~\ref{appmicro}. The eigenstates $\Psi^{JM\pi}$ with energy  $E^{J\pi}$ of the system are  solutions, as usual, of the Schr{\"o}dinger equation associated with the Hamiltonian given in Eq. (\ref{eq1}),
\begin{eqnarray}\label{eq3}
 H\Psi^{JM\pi}=E^{J\pi} \Psi^{JM\pi}.
\end{eqnarray}
The total wave function $\Psi^{JM\pi}$ is a function of the $A-1$ coordinates $\mathbf{r}_{ij}$.

When $A>4$, no exact solutions of Eq.~(\ref{eq3}) can be found and approximate solutions have to be constructed. For those cases, we use a cluster approximation in which  $\Psi^{JM\pi}$ is written in terms of $\alpha$-nucleus wave functions. Because the binding energy of the $\alpha$ particle is large, this approach has been shown to be well adapted to cluster states, and in particular to $^8$Be and $^{12}$C \citep{Ka81,SMO08}. In the particular case of these two nuclei, the wave functions are respectively expressed as
\begin{eqnarray}\label{eq4}
 &&\Psi^{JM\pi}_{^8Be}={\mathcal A} \phi_{\alpha} \phi_{\alpha} g_2^{JM\pi}(\mathbf{\mbox{\boldmath$\rho$}})\nonumber \\
 &&\Psi^{JM\pi}_{^{12}C}={\mathcal A} \phi_{\alpha} \phi_{\alpha}\phi_{\alpha} g_3^{JM\pi}(\mathbf{\mbox{\boldmath $\rho$},R}),
\end{eqnarray}
where $\phi_{\alpha}$ is the $\alpha$ wave function, defined in the $0s$ shell model with an oscillator parameter $b$; ${\mathcal A}$ is the antisymmetrization operator between the $A$ nucleons of the system. For two-cluster systems, the wave function $g_2^{JM\pi}(\mathbf{\mbox{\boldmath $\rho$}})$ depends on the relative coordinate $\mathbf{\mbox{\boldmath$\rho$}}$ between the two $\alpha$ particles. For three-cluster systems, $\mathbf{R}$ is the relative distance between two $\alpha$ particles, and $\mathbf{\mbox{\boldmath $\rho$}}$ the relative coordinate between the third $\alpha$ particle and the $^8$Be center of mass. The relative wave functions, $g_2$ and $g_3$, are obtained by solving the Schr{\"o}dinger equation~(\ref{eq3}).

One then needs to specify the nucleon-nucleon potential $V_N(\mathbf{r}_{ij})$. We shall use the microscopic interaction model \citep{TLT77} which contains one linear parameter (admixture parameter $u$), whose standard value is $u=1$. It can be slightly modified to reproduce important inputs, such as the resonance energy of the Hoyle state. The binding energies of the deuteron ($-2.22$ MeV) and of the $\alpha$ particle ($-24.28$ MeV) do not depend on $u$. For the deuteron, the Schr\"odinger equation is solved exactly. More details about the model are given in Appendix~\ref{appmicro}.

To take into account the variation of the fundamental constants, we introduce the parameters $\delta_\alpha$ and $\delta_{NN}$ to characterize the change of the strength of the electromagnetic and nucleon-nucleon interaction respectively. This is implemented by modifying the interaction potential~(\ref{eq2}) so that
\begin{eqnarray}\label{eq2b}
 V(\mathbf{r}_{ij})=(1+\delta_\alpha)V_C(\mathbf{r}_{ij})+(1+\delta_{NN})V_N(\mathbf{r}_{ij}).
\end{eqnarray}
Such a modification will affect $B_D$ and the energy levels of $^8$Be and $^{12}$C simultaneously.
Of course, one could have imagined a more complex parameterization of the variations (e.g., 
by varying all quantities in Eq. (\ref{vs})), but since we expect to consider only small 
variations of all quantities, as an approximation, the system should be linear in the variations
and our approach should be sufficient for extracting the physical effects of any such small variation.

\subsection{Sensitivity of the nuclear parameters}\label{sens}

For each set of values ($\delta_\alpha,\delta_{NN}$) we solve Eq.~(\ref{eq3}) with the interaction potential~(\ref{eq2b}). We emphasize that the parameter $u$ is determined from the experimental $^8$Be and $^{12}$C$(0^+_2)$ energies ($u=0.954$). We assume that $\delta_{NN}$ varies in the range $[-0.015,0.015]$.

\begin{figure}
\centering
 \resizebox{\hsize}{!}{\includegraphics{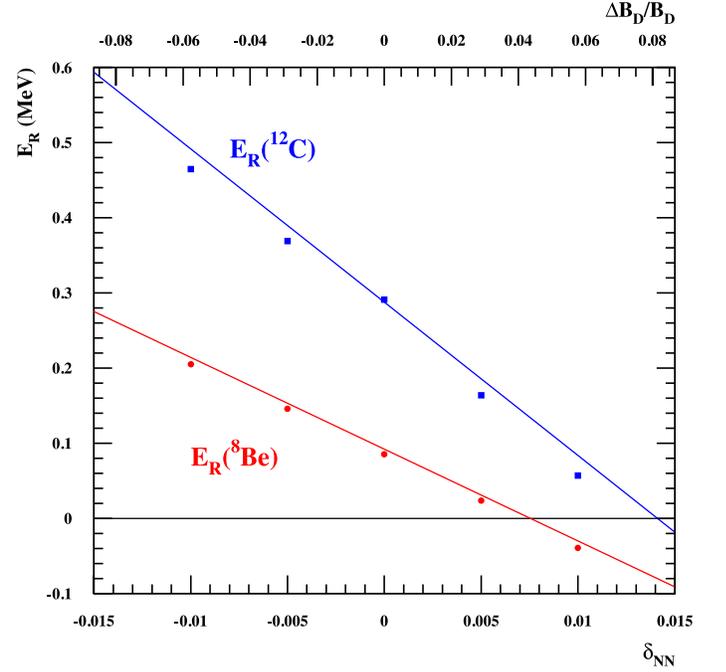}}
\caption{Variation of the resonance energies as a function of $\delta_{NN}$. The symbols represent the results of the microscopic calculation while the lines correspond to the adopted linear relationship between $E_R$ and $\delta_{NN}$.}
\label{fener}
\end{figure}

First, concerning the deuteron, this analysis implies that its binding energy scales as
\begin{eqnarray}\label{BDdNN}
  {\Delta}B_D/B_D = 5.716\times\delta_{NN} \, .
\end{eqnarray}
Second, concerning $^8$Be and $^{12}$C, we can extract the sensitivity of $E_R$($^8$Be) and $E_R$($^{12}$C). They scale as
\begin{eqnarray}
E_R(^8{\rm Be}) = \left( 0.09184-12.208\times\delta_{NN}\right) \, \mathrm{MeV}
\end{eqnarray}
and 
\begin{eqnarray}
E_R(^{12}{\rm C}) =  \left(0.2876-20.412\times\delta_{NN}\right) \ \mathrm{MeV}.
\end{eqnarray}
The numerical results for the sensitivities of $E_R(^8{\rm Be})$ and $E_R(^{12}{\rm C})$ to $\delta_{NN}$ as well as the above linear fits are shown in Fig.~\ref{fener}. The effect of $\delta_\alpha$ on these quantities is negligible. Note that for $\delta_{NN}\gtrsim 0.007$,  $E_R(^8${\rm Be}) is negative and $^8$Be would become stable. Using the bijective relation~(\ref{BDdNN}) between $B_D$ and $\delta_{NN}$ we can also express our results as
\begin{eqnarray}
 E_R(^8{\rm Be})&=& \left( 0.09184- 2.136\times\Delta B_D/B_D\right)  \ \mathrm{MeV}, \\
 E_R(^{12}{\rm C}) &=&  \left(0.2876- 3.570 \times \Delta B_D/B_D\right) \ \mathrm {MeV}.
\end{eqnarray}
It follows that  the energy of the Hoyle level with respect to the $3 \alpha$ threshold (and not with respect to $^8$Be+$\alpha$ threshold) is given by (see Eq.~(\ref{Q3alpha}))
\begin{eqnarray}\label{Qdelta}
 Q_{\alpha\alpha\alpha} &=&\left(0.37945- 5.706 \times\Delta B_D/B_D
 \right) \ \mathrm{MeV}, \\
 &=&\left(0.37945- 32.620\times \delta_{NN}\right) \ \mathrm{MeV}.
\end{eqnarray}

To estimate the effect of $\delta_\alpha$ in Eq. (7), we can approximate the Coulomb energy by $(3/5) Z(Z-1) \alpha_{em} \hbar c / R_c$ where $R_c = 1.3A^{1/3}$ fm which gives 9 MeV for $^{12}$C and 0.9 MeV for $^{4}$He. The variation of $Q_{\alpha\alpha\alpha}$ is thus of the order of $+6$ MeV $\times~\delta_\alpha$. The direct effect of ${\delta_\alpha}$ is  thus of opposite sign but considerably less important.
This is in qualitative agreement with \citet{ober2000} and \citet{ober2001}.

\subsection{Sensitivity of the $3\alpha$-reaction rate}

The method described above provides a consistent way to evaluate the sensitivity of the $3\alpha$-reaction rate to a variation of the constants. This rate has been computed numerically as explained in \citet{NACRE} and as described in Appendix~\ref{s:rate} where both an analytical approximation valid for sharp resonances and a numerical integration are performed.

\begin{figure}
\centering
 \resizebox{\hsize}{!}{\includegraphics{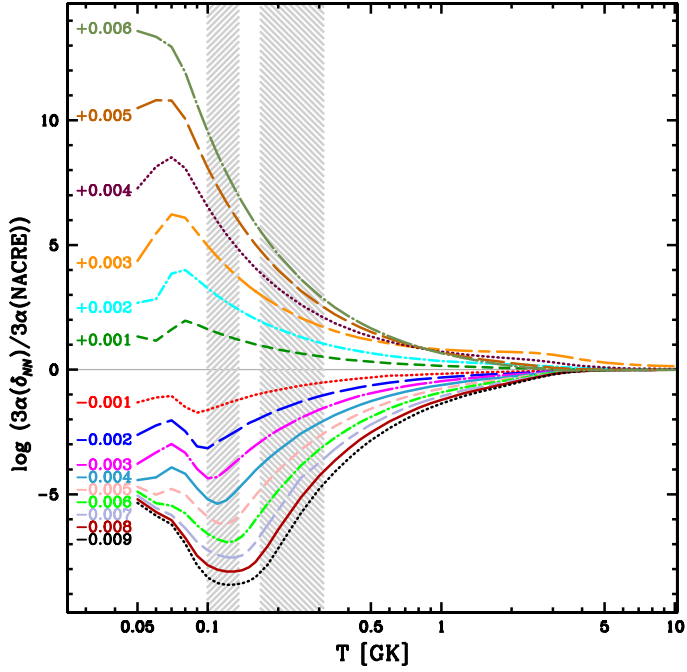}}
\caption{The ratio between the 3$\alpha$ rate obtained for $-0.009 \leq \delta_{NN} \leq +0.006$ and the NACRE rate, as a function of temperature. Hatched areas: values of $T_{\rm c}$ where H  and He burning phases take place in a 15 $M_{\sun}$  model at $Z=0$.}
\label{fratio}
\end{figure}

The variation of the partial widths of both reactions have been computed in Appendix~B and are depicted in Fig.~\ref{fwidths}. Together with the results of the previous section and the details of the Appendix~B, we can compute the $3\alpha$-reaction rate as a function of temperature and $\delta_{NN}$. This is summarized in Fig. \ref{fratio} which compares the rate for different values of $\delta_{NN}$ to the NACRE rate \citep{NACRE}, which is our reference when no variation of constants is assumed (i.e. $\delta_{NN}=0$). One can also refer to Fig.~\ref{frates} which compares the full numerical integration to the analytical estimation~(\ref{eq:anal}) which turns out to be excellent in the range of temperatures of interest. As one can see, for positive values of $\delta_{NN}$, the resonance energies are lower, so that the $3\alpha$ process is more efficient (see Appendix \ref{s:rate}).

Let us compare the result of Fig.~\ref{fratio}, which gives $y\equiv \log[\lambda_{3\alpha}(\delta_{NN})/\lambda_{3\alpha}(0)]$  to a simple estimate. Using the analytic expression~(\ref{eq:anal}) for the reaction rate,  valid only for a sharp resonance, $y$ is simply given by
\begin{equation}
 y =\frac{1}{\ln 10} s_{\delta_{NN}}
\end{equation}
where the sensitivity $s_{\delta_{NN}}\equiv {{\rm d} \ln\lambda_{3\alpha} }/{{\rm d}\ln\delta_{NN}}$
is given, from Eq.~(\ref{Qdelta}), by $s_{\delta_{NN}}= \delta_{NN}\times(32.62{\rm MeV})/kT$. We conclude that
\begin{equation}
 y = 1.644 \times \left(\frac{\delta_{NN}}{10^{-3}}\right)\left(\frac{T}{10^8\,{\rm K}}\right)^{-1}.
\end{equation}
This gives the correct order of magnitude of the curves depicted in Fig.~\ref{fratio} as well as their scalings with $\delta_{NN}$ and with temperature, as long as $T_9 > 0.1$. At lower temperatures differences arise from the fact that the analytical expression for the reaction rate is no longer accurate (see Appendix~\ref{s:rate}).

The sensitivity to a variation of the intensity of the N-N interaction arises from the fact that $ {{\rm d} Q_{\alpha\alpha\alpha} }/{{\rm d}\delta_{NN}}\sim 10^2 Q_{\alpha\alpha\alpha}$.  The fact that the typical correction to the resonant energies is of order 10 MeV ($\times~\delta_{NN}$) compared to the resonant energies themselves which are of order 0.1 MeV, allows one to put relatively strong constraints on any variation. This is reminiscent of the case of the resonance producing an excited state of $^{150}$Sm of importance in setting constraints on the variation of couplings using the Oklo reactor \citep{oklo0,oklo1,Fujii:1998kn,oklo2,oklo3}. In that case, the resonant energy is 0.1 eV compared to corrections of order 1 MeV due to changes in the fine structure constant, leading to limits on $\Delta \alpha_{em}/ \alpha_{em}$ of order $10^{-7}$. 

\subsection{Using the Deuterium binding energy as a link to fundamental constants}

The nuclear model described above introduces the parameter $\delta_{NN}$ which is itself not directly related to a set of fundamental constants such as gauge and Yukawa couplings. In order to make such a connection, we make use of previous analyses relating the deuterium binding energy $B_D$ to fundamental constants.

Using a potential model, the dependence of $B_D$ on the nucleon, $\sigma$-meson and $\omega$-meson has been estimated~\citep{bbn2,sig3,sig1,sig2,coc07,dd}. Furthermore, using the quark matrix elements for the nucleon, variations in $B_D$ can be related to variations in the light quark masses (particularly the strange quark) and thus to the corresponding quark Yukawa couplings and Higgs vev, $v$. The remaining sensitivity of $B_D$ to a dimensionful quantity is ascribed to the QCD scale $\Lambda$. In \citet{coc07}, it was concluded that
\begin{equation}
\frac{\Delta B_D}{B_D} = 18\frac{\Delta\Lambda}{\Lambda} -
17\left(\frac{\Delta v}{v}+\frac{\Delta h_s}{h_s} \right).
\end{equation}
Eq.~(\ref{BDdNN}) can then link any constraint on $\delta_{NN}$ to the three fundamental constants $(h_s,v,\Lambda)$. 

Further relations are possible in the context of unified theories of gauge interactions. From the low energy expression for $\Lambda_{QCD}$,
\begin{equation}
\Lambda = \mu \left(\frac{m_c \, m_b \, m_t}{\mu^3} \right)^{2/27} \, \exp\left(-\frac{2\pi}{9\alpha_s(\mu)} \right) \,,
\end{equation}
one can determine the relation between the changes in  $\Lambda$ and the gauge couplings and quark masses  \citep{campolive95,lang,dent,cal,dam02},
\begin{eqnarray}
\label{DeltaLambda}
\frac{\Delta \Lambda}{\Lambda} &= &R \, \frac{\Delta \alpha_{em}}{\alpha_{em}} +
\frac{2}{27} \left(3 \, \frac{\Delta v}{v} + \frac{\Delta h_c}{h_c} + \frac{\Delta h_b}{h_b}
+  \frac{\Delta h_t}{h_t} \right) \,.
\end{eqnarray}
Typical values for $R$ are of order 30 in many grand unified theories, but there is considerable model dependence in this coefficient \citep{dine}.

Furthermore, in theories in which the electroweak scale is derived by dimensional transmutation, changes in the Yukawa couplings (particularly the top Yukawa) leads to exponentially large changes in the Higgs vev. In such theories, 
\begin{equation}
\frac{\Delta v }{v} \sim S \frac{\Delta h}{h}
\label{enhance2}
\end{equation}
with $S \sim 160$, though there is considerable model dependence in this value as well.

Finally, using the relations in Eqs. (\ref{DeltaLambda}) and (\ref{enhance2}), we can write 
\begin{eqnarray}
\label{DeltaBd3}
 \frac{\Delta B_D}{B_D} &=& -13 (1+S) \, \frac{\Delta h}{h}
            + 18R \, \frac{\Delta \alpha_{em}}{\alpha_{em}} \, .
\end{eqnarray}
If in addition, we relate the gauge and Yukawa couplings through $\Delta h/h = (1/2)\Delta \alpha_{em}/\alpha_{em}$, we can further write, 
\begin{eqnarray}
\label{DeltaBd4}
 \frac{\Delta B_D}{B_D} &=& -[6.5(1+S)-18R] \frac{\Delta \alpha_{em}}{\alpha_{em}} \, .
\end{eqnarray}

An alternative investigation \citep{pi1,pi2,pi3} suggests a large dependence of $B_D$ on the pion mass $m_\pi$,
\begin{equation}
 \frac{\Delta B_D}{B_D} = -r \frac{\Delta m_\pi}{m_\pi}
\end{equation}
where $r$ is expected to range between 6 and 10. Again, this allows one to related $B_D$, and thus $\delta_{NN}$ to $(h,v,\alpha_{em})$.

As these two examples demonstrate, the main problem arises from the difficulty to determine the role of the QCD parameter in low energy nuclear physics. They show, however, that such a link can be drawn, even though it is strongly model-dependent. 

\section{Stellar implications}\label{sec_stellar}

The Geneva stellar code was adapted to take into account the reaction rates computed above. The version of the code we use is the one described in \citet{emchm08}. Here, we only consider models of 15 $M_{\sun}$ and 60 $M_{\sun}$ without rotation and assume an initial chemical composition given by $X= 0.7514$, $Y= 0.2486$ and $Z=0$. This corresponds to the BBN abundance of He at the baryon density determined by WMAP \citep{wmap5y} and at zero metallicity as is expected to be appropriate for Population III stars. For 16 values of the free parameter $\delta_{NN}$ in the range $-0.009\leq \delta_{NN} \leq +0.006$, we computed a stellar model which was followed up to the end of core He burning (CHeB). As we will see, beyond this range in $\delta_{NN}$, stellar nucleosynthesis is unacceptably altered. Note that for some of the most extreme cases, the set of nuclear reactions now implemented in the code should probably be adapted for a computation of  the advanced evolutionary phases. 

Focusing on the limited range in $\delta_{NN}$ will allow us to study the impact of a change of the fundamental constants on the production of carbon and oxygen in Pop III massive stars. In this context, we recall that the observations of the most iron poor stars in the halo offer a wonderful tool to probe the nucleosynthetic impact of the first massive stars in the Universe. Indeed these halo stars are believed to form from material enriched by the ejecta of the first stellar generations in the Universe. Their surface chemical composition (at least on the Main Sequence), still bear the mark of the chemical composition of the cloud from which they formed and thus allow us to probe  the nucleosynthetic signature of the first stellar generations. Any variation of the fundamental constants which for instance would prevent the synthesis of carbon and/or  oxygen would be very hard to conciliate with present day observations of the most iron poor stars. For instance the two most iron poor stars \citep{Christlieb2004, frebel08} both show strong overabundances of carbon and oxygen with respect to iron.

Our results for 15 $M_{\sun}$ and 60 $M_{\sun}$ stars are presented in \S~\ref{subsec_stellar_a} and \S~\ref{subsec_stellar_b} respectively.

\subsection{15 $M_{\sun}$ mass star}\label{subsec_stellar_a}

\begin{figure*}
\centering
 \resizebox{\hsize}{!}{\includegraphics{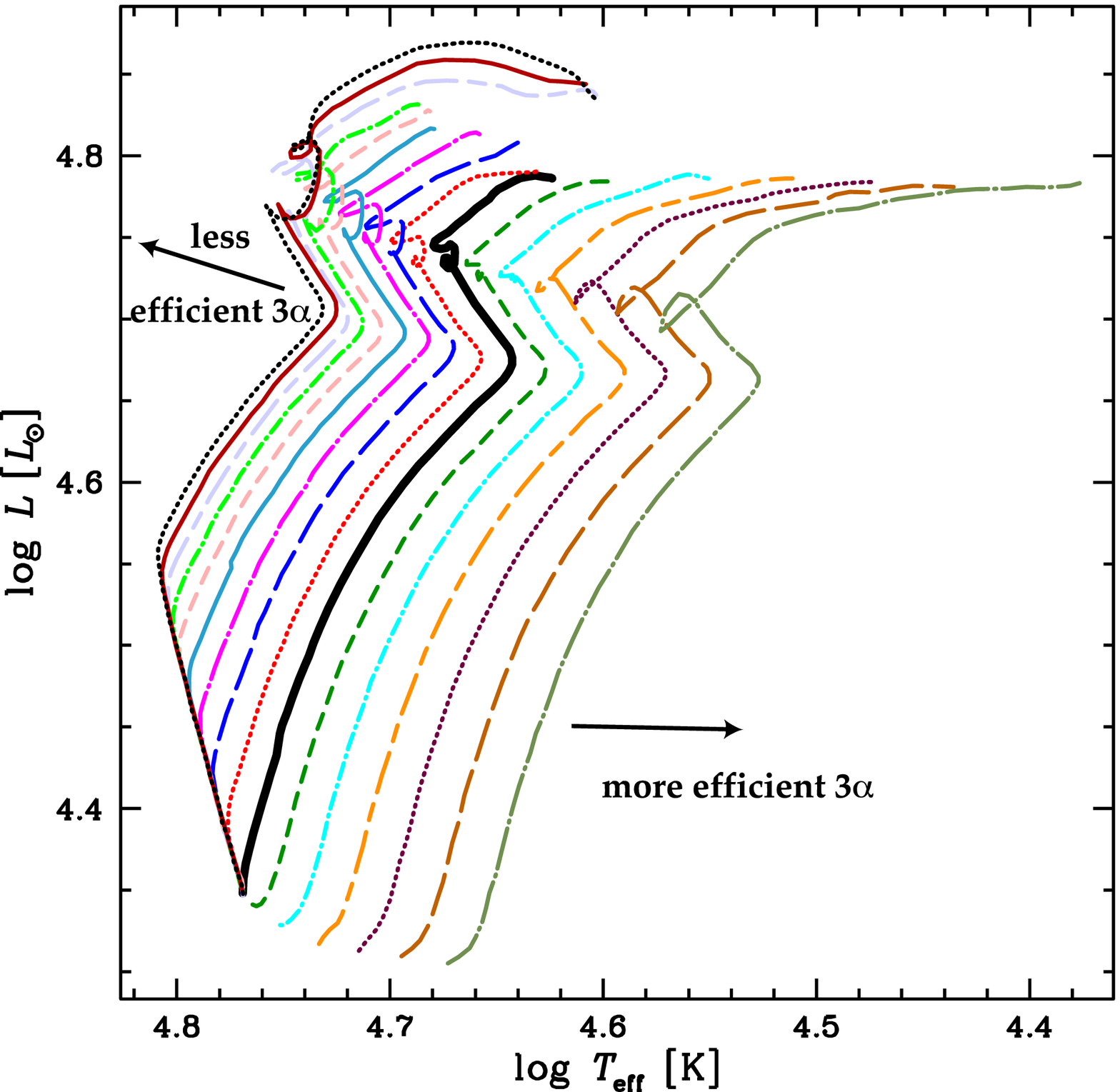}\hspace{1cm}\includegraphics{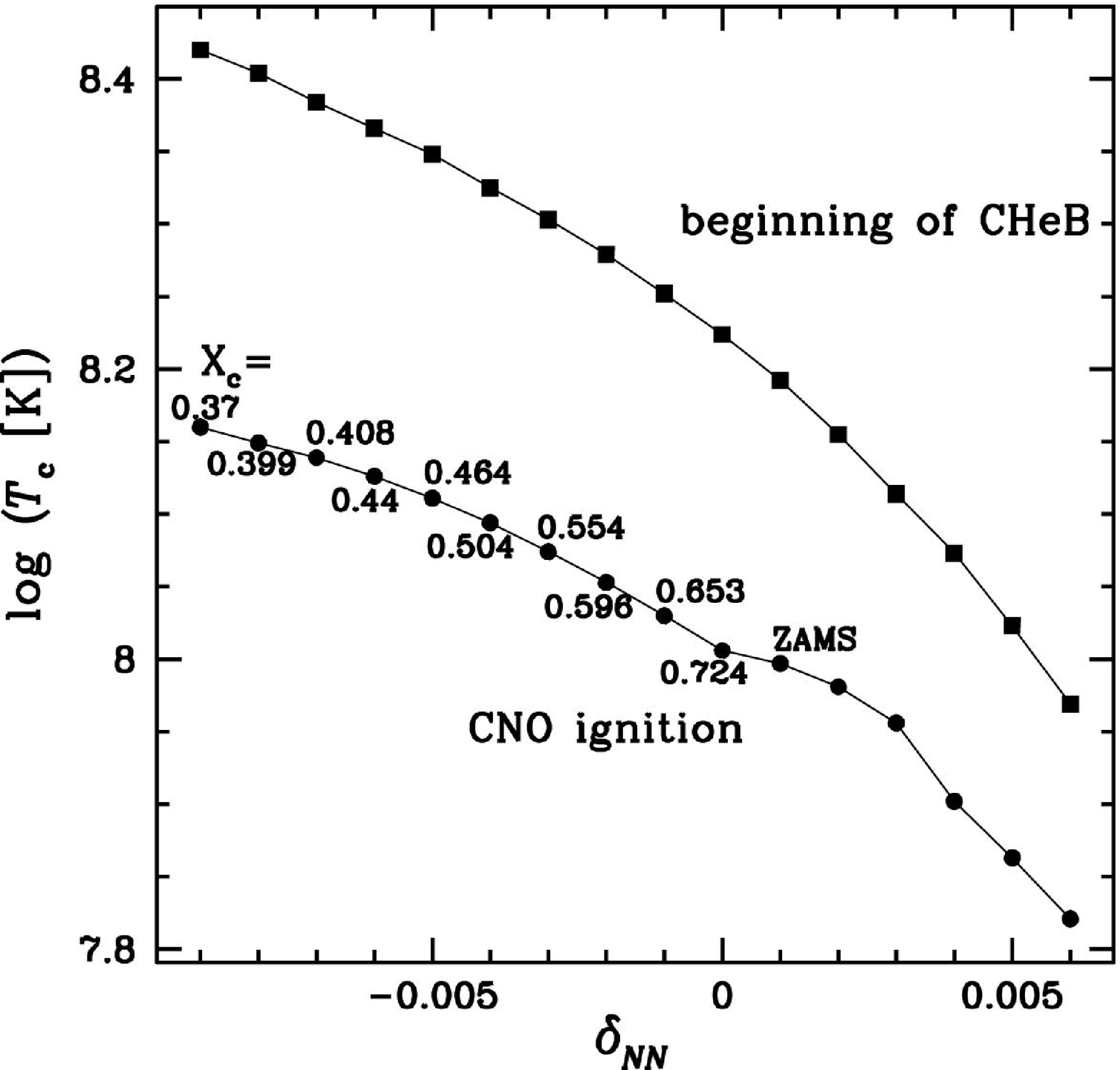}}
\caption{\emph{Left panel:} HR diagrams for 15 $M_{\sun}$  models with $\delta_{NN}=0$ (thick black), and ranges from left to right from $-0.009$ to $+0.006$ in steps of 0.001 (using same colour code as Fig.~\ref{fratio}). \emph{Right panel:} The central temperature at CNO ignition (circles) and at the beginning of CHeB (squares) as a function of $\delta_{NN}$. The labels on the CNO-ignition curve shows the central H mass fraction at that moment. Note that above $\delta_{NN} \sim +0.001$, CNO ignition occurs on or before the ZAMS.}
\label{fdhrtc15}
\end{figure*}

Figure~\ref{fdhrtc15} (\emph{left}) shows the HR diagram for the models with $\delta_{NN}$ between -0.009 et +0.006 in increments of 0.001 (from left to right). Once the CNO cycle has been triggered (see below), the Main Sequence (MS) tracks are shifted towards cooler $T_{\rm eff}$ for increasing $\delta_{NN}$. There is a difference of about 0.20 dex between the two extreme models. Figure~\ref{fdhrtc15} (\emph{right}) shows the central temperature at the moment of the CNO-cycle ignition (lower curve). On the zero-age main sequence (ZAMS), the standard ($\delta_{NN} = 0$) model has not yet produced enough $^{12}$C to be able to rely on the CNO cycle, so it starts by continuing its initial contraction until the CNO cycle ignites. In this model, CNO ignition occurs when the central H mass fraction reaches 0.724, \textsl{i.e.} when less than 3\% of the initial H has been burned. Models with $\delta_{NN} < 0$ (\textsl{i.e.} a lower 3$\alpha$ rate) yield a phase of contraction which is longer for lower $\delta_{NN}$ (i.e. larger $|\delta_{NN}|$): in these models, the less efficient 3$\alpha$ rates need a higher $T_{\rm c}$ to produce enough $^{12}$C for triggering the CNO cycle. Models with $\delta_{NN} > 0$ (\textsl{i.e.} a higher 3$\alpha$ rate) are directly sustained by the CNO cycle on the ZAMS: the star can more easily counteract its own gravity and the initial contraction is stopped earlier, so H burning occurs at lower $T_{\rm c}$ and $\rho_{\rm c}$ (Fig.~\ref{fdhrtc15}, \textsl{right}), \textsl{i.e.} at a slower pace. The MS lifetime, $\tau_{\rm MS}$, is sensitive to the pace at which H is burned, so it increases with $\delta_{NN}$. The relative difference between the standard model MS lifetime $\tau_{\rm MS}$ at $\delta_{NN} = 0$ and $\tau_{\rm MS}$ at $\delta_{NN} =-0.009$ (+0.006) amounts to -17\% (+19\%).

\begin{figure}
\centering
 \resizebox{\hsize}{!}{\includegraphics{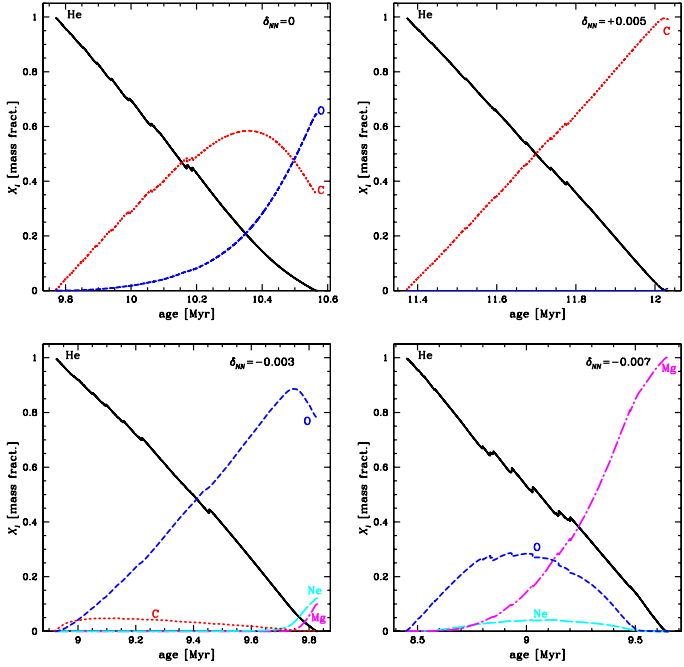}}
\caption{The evolution of the central mass fraction for the main chemical species inside the core of the 15 $M_{\sun}$ models during core He burning. {\it Top left:} standard case (representative of case I, see text), {\it bottom left:} $\delta_{NN} = -0.003$ (case II), {\it bottom right:} $\delta_{NN} = -0.007$ (case III), {\it top right:} $\delta_{NN} = +0.005$ (case IV).}
\label{fxco1}
\end{figure}

\begin{figure}
\centering
 \resizebox{.8\hsize}{!}{\includegraphics{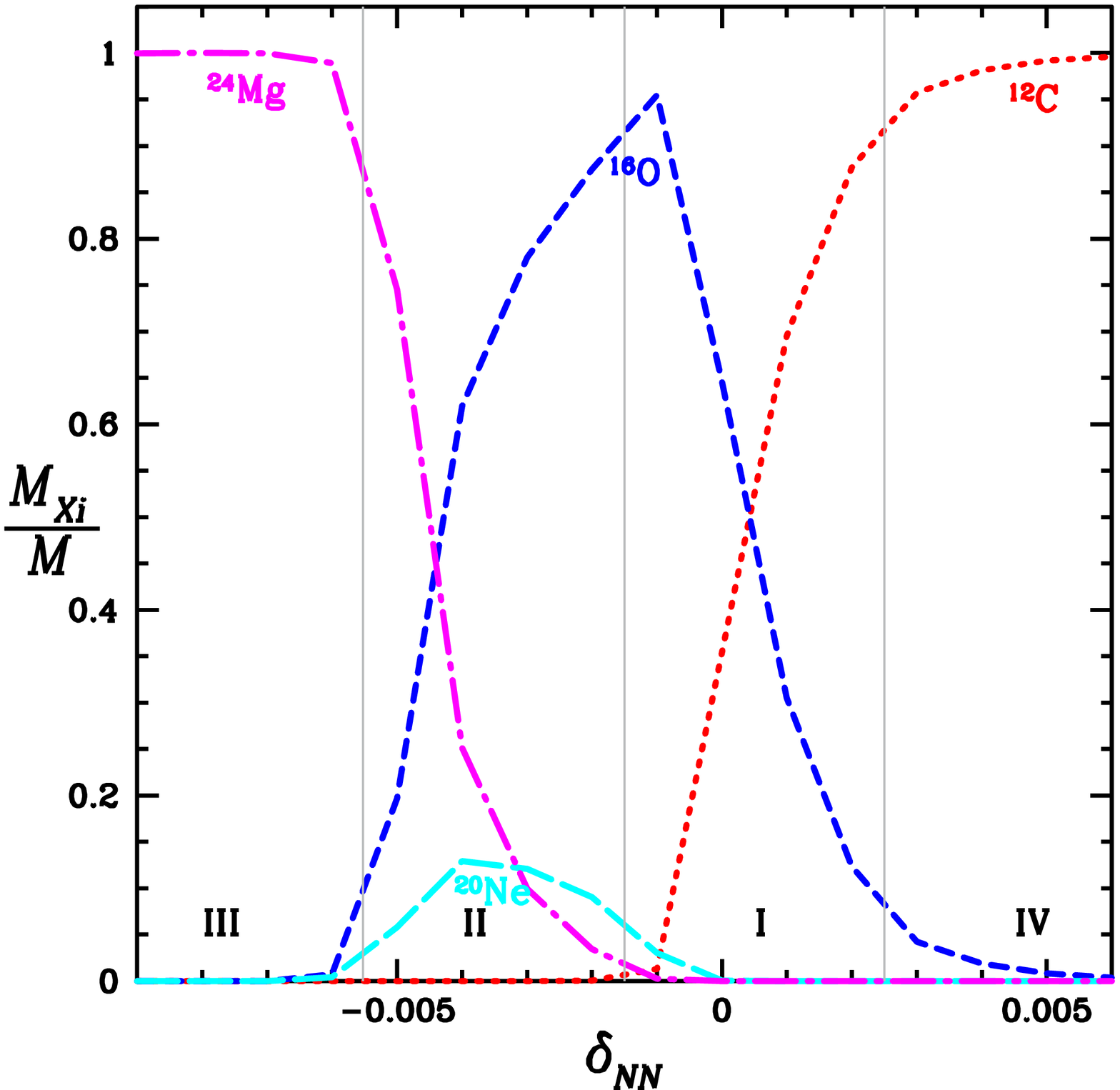}}
\caption{The composition of the core at the end of the central He burning in the 15 $M_{\sun}$ models as a function of $\delta_{NN}$.}
\label{sum15}
\end{figure}

While the differences in the 3$\alpha$ rates do not lead to strong effects in the evolution characteristics
on the MS, the CHeB phase amplifies the differences between the models. The upper curve of Fig.~\ref{fdhrtc15} (\emph{right}) shows the central temperature at the beginning of CHeB. There is a factor of 2.8 in temperature between the models with $\delta_{NN} = -0.009$ and $+0.006$. To get an idea of what this difference represents, we can relate these temperatures to the grid of Pop III models computed by \citet{mar01}. The 15 $M_{\sun}$ model with $\delta_{NN} = -0.009$ starts its CHeB at a higher temperature than a standard 100 $M_{\sun}$ of the same stage. In contrast, the model with $\delta_{NN} = +0.006$ starts its CHeB phase with a lower temperature than a standard 12 $M_{\sun}$ star at CNO ignition. Table~\ref{tmod1} presents the characteristics of the models for each value of $\delta_{NN}$ at the end of CHeB. From these characteristics, we distinguish four different cases (see the last column of Table~\ref{tmod1} and Fig.~\ref{fxco1}):
\begin{itemize}
\item[I] In the standard model and when $\delta_{NN}$ is very close to 0,  $^{12}$C is produced during He burning until the central temperature is high enough for the $^{12}$C($\alpha$,$\gamma$)$^{16}$O reaction to become efficient: during the last part of the CHeB phase, the $^{12}$C is processed into $^{16}$O. The star ends its CHeB phase with a core composed of a mixture of $^{12}$C and $^{16}$O (see the top left panel of Fig.~\ref{fxco1}). 
\item[II] If the 3$\alpha$ rate is weakened ($-0.005 \le \delta_{NN} \le -0.002$), $^{12}$C is produced at a slower pace, and $T_{\rm c}$ is high from the beginning of the CHeB phase, so the $^{12}$C($\alpha$,$\gamma$)$^{16}$O reaction becomes efficient very early: as soon as some $^{12}$C is produced, it is immediately transformed into $^{16}$O. The star ends its CHeB phase with a core composed mainly of $^{16}$O, without any $^{12}$C and with an increasing fraction of $^{24}$Mg for decreasing $\delta_{NN}$ (see the bottom left panel of Fig.~\ref{fxco1}).
\item[III] For still weaker 3$\alpha$ rates ($\delta_{NN} \le -0.006$), the central temperature during CHeB is such that the $^{16}$O($\alpha$,$\gamma$)$^{20}$Ne($\alpha$,$\gamma$)$^{24}$Mg chain becomes efficient, reducing the final $^{16}$O abundance. The star ends its CHeB phase with a core composed of nearly pure $^{24}$Mg (see the bottom right panel of Fig.~\ref{fxco1}). Because the abundances of both carbon and oxygen are completely negligible, we do not list the irrelevant value of C/O for these cases.
\item[IV] If the 3$\alpha$ rate is strong ($\delta_{NN} \ge +0.003$), $^{12}$C is very rapidly produced, but $T_{\rm c}$ is so low that the $^{12}$C($\alpha$,$\gamma$)$^{16}$O reaction can hardly enter into play: $^{12}$C is not transformed into $^{16}$O. The star ends its CHeB phase with a core almost purely composed of $^{12}$C (see the top right panel of Fig.~\ref{fxco1}).
\end{itemize}
These results are summarized in Fig.~\ref{sum15} which shows the composition of the core at the end of the CHeB phase. One can clearly see the dramatic change in the core composition as a function of $\delta_{NN}$ showing a nearly pure Mg core at large and negative $\delta_{NN}$, a dominantly O core at low but negative $\delta_{NN}$, and a nearly pure C core at large and positive $\delta_{NN}$. These results are qualitatively consistent with those found by \citet{schlattl2004} for Population I type stars. Note that their cases with $\Delta E_{\rm R}=\pm100$ keV correspond roughly to our $\delta_{NN}\approx\mp0.005$.

\begin{table}
 \begin{minipage}{.5\textwidth}
\centering
 \renewcommand{\footnoterule}{}
\caption{Characteristics of the 15 $M_{\sun}$ models with $\delta_{NN}$ ranging from -0.009 to +0.006 at the end of core He burning. The MS lifetime, $\tau_{\rm MS}$, and the core He-burning duration, $\tau_{\rm CHeB}$, are expressed in Myr, the CO-core mass, $M_{\rm CO}$, is in $M_{\sun}$,  $X({\rm C})$ is the central value for the carbon mass fraction and C/O is the ratio of the carbon to oxygen mass fractions.}
\begin{tabular}{crrrccc}
\hline \hline
$\delta_{NN}$ & \multicolumn{1}{c}{$\tau_{\rm MS}$} & \multicolumn{1}{c}{$\tau_{\rm CHeB}$} & \multicolumn{1}{c}{$M_{\rm CO}$\footnote{mass coordinate where the abundance of $^4$He drops below 10$^{-3}$}} & \multicolumn{1}{c}{$X({\rm C})$\footnote{Here, $X({\rm C}) = X(^{12}C)$}} & \multicolumn{1}{c}{C/O} & case \\
\hline
 -0.009 & 8.224 & 1.344 & 3.84 & 4.4e-10 & -- & III \\
 -0.008 & 8.285 & 1.276 & 3.83 & 2.9e-10 & -- &  \\
 -0.007 & 8.308 & 1.200 & 3.38 & 8.5e-10 & -- &  \\
 -0.006 & 8.401 & 1.168 & 3.61 & 4.2e-07 & -- & \\
  \hline
-0.005 & 8.480 & 1.130 & 3.59 & 5.9e-06 & 3.0e-05 & II \\
 -0.004 & 8.672 & 0.933 & 3.60 & 3.2e-05 & 5.2e-05 & \\
 -0.003 & 8.790 & 0.905 & 3.60 & 1.3e-04 & 1.7e-04 & \\
 -0.002 & 9.046 & 0.892 & 3.61 & 5.6e-04 & 6.4e-04 & \\
 \hline
 -0.001 & 9.196 & 0.888 & 3.70 & 0.013 & 0.014 & I \\
 0          & 9.640 & 0.802 & 3.65 & 0.355 & 0.550 & \\
+0.001 & 9.937 & 0.720 & 3.61 & 0.695 & 2.278 & \\
+0.002 & 10.312 & 0.684 & 3.62 & 0.877 & 7.112 & \\
 \hline
+0.003 & 10.677 & 0.664 & 3.62 & 0.958 & 22.57 & IV \\
+0.004 & 10.981 & 0.659 & 3.62 & 0.981 & 52.43 & \\
+0.005 & 11.241 & 0.660 & 3.61 & 0.992 & 123.9 & \\
+0.006 & 11.447 & 0.661 & 3.55 & 0.996 & 270.2 & \\
\hline
\end{tabular}
\label{tmod1}
 \end{minipage}
\end{table}%

Table~\ref{tmod1} shows also the core size at the end of CHeB. As in \citet{hlw00} the mass, $M_{\rm CO}$, is determined as the mass coordinate where the mass fraction of $^4$He drops below 10$^{-3}$. The mass of the CO core increases with decreasing $\delta_{NN}$, the increase amounting to 8\% between $\delta_{NN}=+0.006$ and -0.009. This effect is due to the higher central temperature and greater compactness at low $\delta_{NN}$. The same effect was found by other authors \citep{schlattl2004,tur}. As shown by these authors, this effect is expected to have an impact on the remnant mass and thus on the strength of the final explosion.

\subsection{$60 M_{\sun}$ mass star}\label{subsec_stellar_b}

As it is widely believed that Pop III stars are massive, we next present results for 60 $M_{\sun}$ models (at $Z=0$). The characteristics of these models for different values of $\delta_{NN}$ are collected in Table~\ref{tmod60}.

\begin{table}
\centering
\caption{Characteristics of the 60 $M_{\sun}$ models with $\delta_{NN}$ ranging from -0.009 to +0.006. $\tau_{\rm MS}$ and $\tau_{\rm CHeB}$ are expressed in Myr, $M_{\rm CO}$ is in $M_{\sun}$, $X(C)$ is in mass fraction. The mark ``$75$'' after $M_{\rm CO}$ indicates that it has been calculated as the mass coordinate where the C+O abundance rises above 0.75. The mark ``$75$Mg'' after $M_{\rm CO}$ indicates that it is the $^{24}$Mg abundance which has become higher than 0.75.}
\begin{tabular}{crrlccc}
\hline \hline
$\delta_{NN}$ & \multicolumn{1}{c}{$\tau_{\rm MS}$} & \multicolumn{1}{c}{$\tau_{\rm CHeB}$} & \multicolumn{1}{c}{$M_{\rm CO}$} & \multicolumn{1}{c}{$X({\rm C})$} & \multicolumn{1}{c}{C/O} & case \\
\hline
    -0.009 & 3.061 & 0.374 & 22.31$^{75 {\rm Mg}}$ & 2.3e-08 & -- & III \\
    -0.008 & 3.098 & 0.373 & 22.16$^{75 {\rm Mg}}$ & 2.1e-08 & -- &  \\
    -0.007 & 3.133 & 0.387 & 22.43$^{75 {\rm Mg}}$ & 2.4e-08 &  -- &  \\
    -0.006 & 3.171 & 0.418 & 19.71$^{75 {\rm Mg}}$ & 4.3e-05 & -- &  \\
    -0.005 & 3.211 & 0.389 & 22.66$^{75 {\rm Mg}}$ & 1.4e-08 & -- &  \\
   \hline
    -0.004 & 3.252 & 0.382 & 21.90 & 3.5e-05 & 2.7e-04 & II \\
    -0.003 & 3.294 & 0.347 & 21.76 & 2.1e-04 & 4.8e-04 &  \\
    -0.002 & 3.338 & 0.328 & 22.27 & 9.3e-04 & 1.4e-03 &  \\
   \hline
    -0.001 & 3.379 & 0.347 & 18.40$^{75}$ & 0.008 & 0.009 & I \\
    0         & 3.418 & 0.299 & 21.37 & 0.163 & 0.200 &  \\
    +0.001 & 3.458 & 0.267 & 21.59 & 0.513 & 1.062 &  \\
    +0.002 & 3.495 & 0.244 & 21.14 & 0.761 & 3.193 &  \\
    +0.003 & 3.534 & 0.259 & 21.14 & 0.899 & 9.233 &  \\
   \hline
    +0.004 & 3.571 & 0.265 & 19.86$^{75}$ & 0.952 & 19.89 & IV \\
    +0.005 & 3.607 & 0.808 & 27.22 & 0.963 & 32.85 &  \\
    +0.006 & 3.644 & -- & -- & -- & -- &  \\
\hline
\end{tabular}
\label{tmod60}
\end{table}

\begin{figure*}
\centering
 \resizebox{\hsize}{!}{\includegraphics{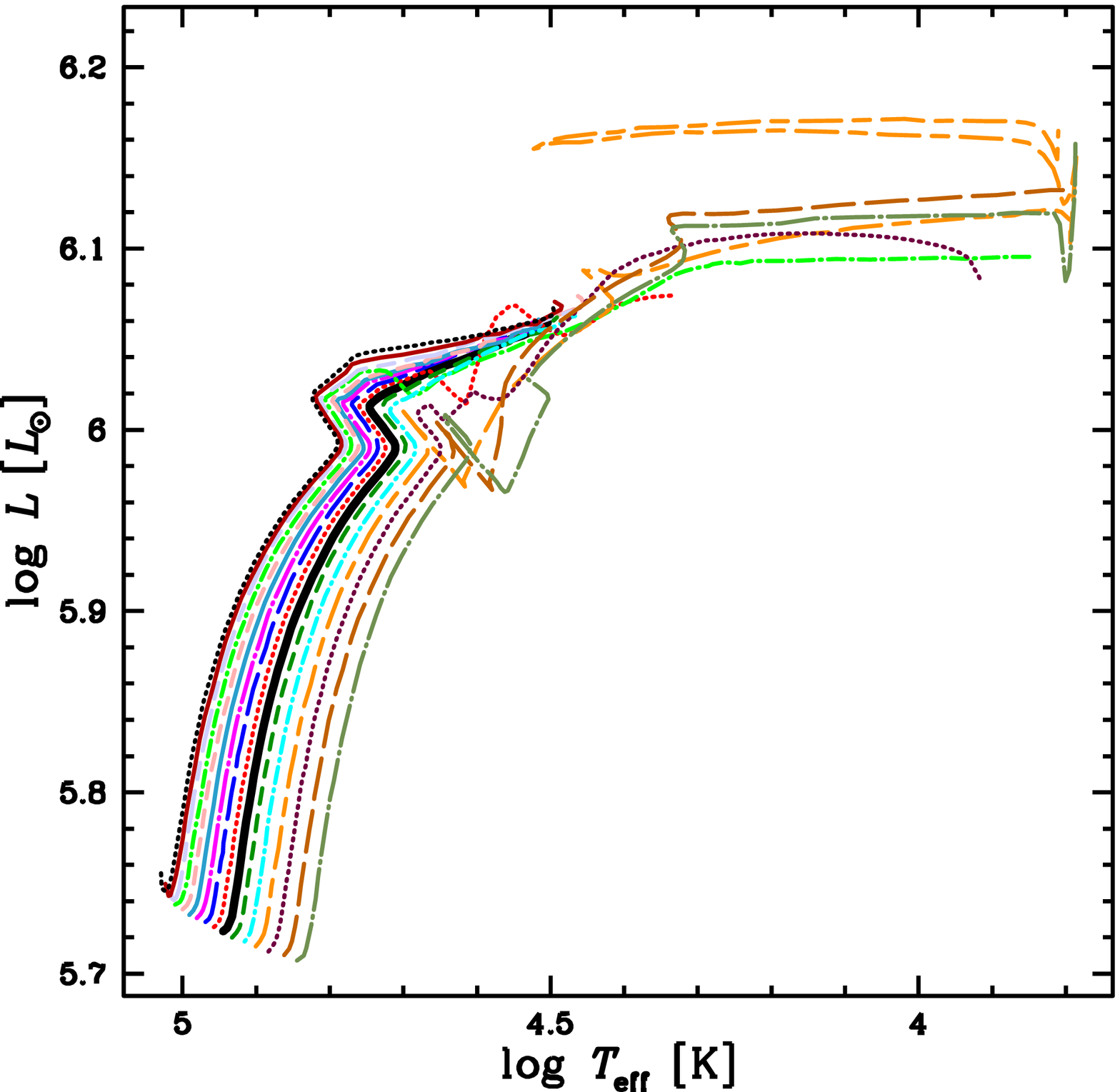}\hspace{1cm}\includegraphics{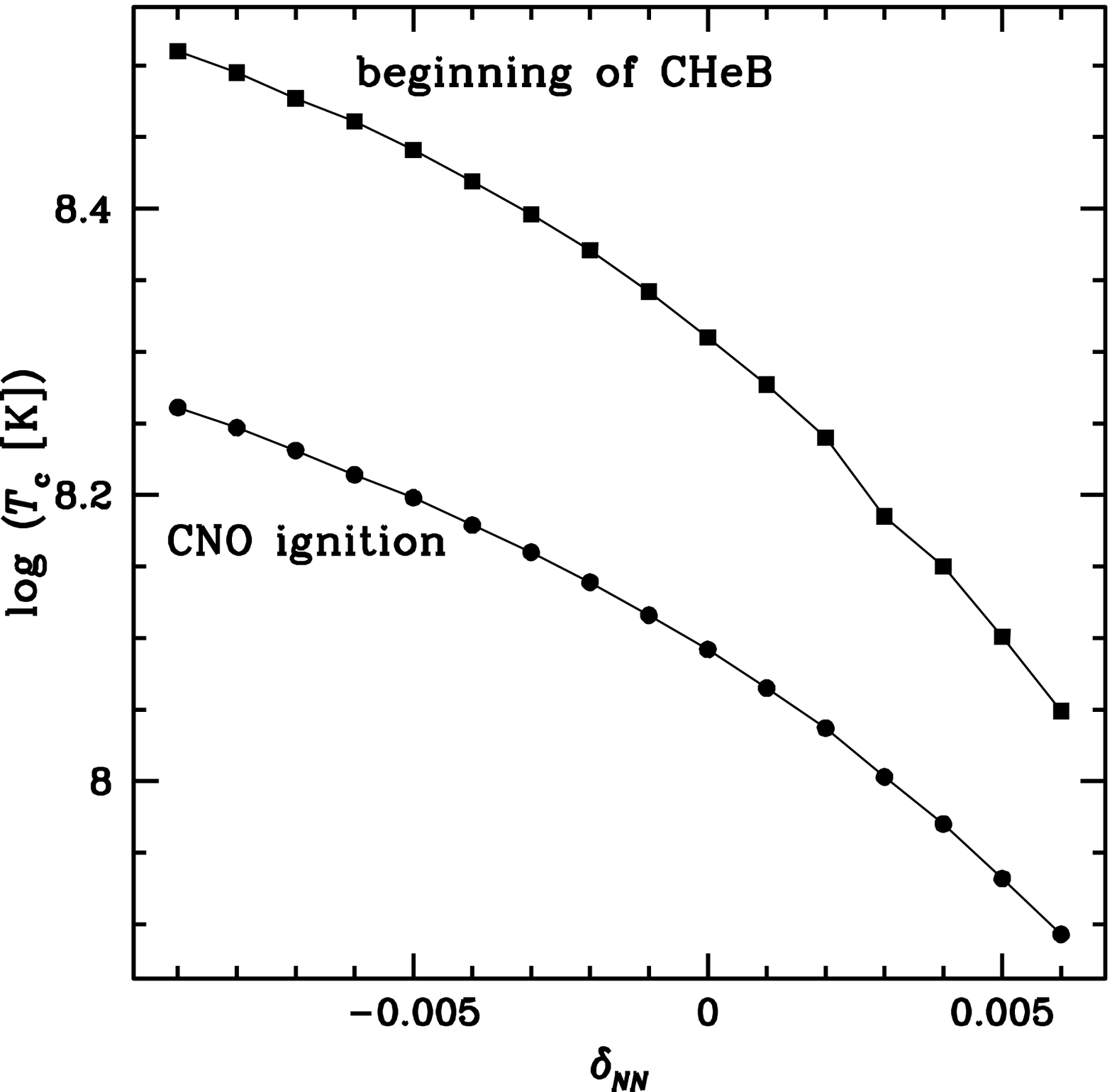}}
\caption{\emph{Left panel:} HR diagrams for 60 $M_{\sun}$  models with $\delta_{NN}=0$ (thick black), and ranges from left to right from $-0.009$ to $+0.006$ in steps of 0.001 (using the same colour code as in Figs.~\ref{fratio} and ~\ref{fdhrtc15}). \emph{Right panel:} The central temperature at CNO ignition (circles) and at the beginning of core He burning (squares) as a function of $\delta_{NN}$. In the 60 M$_{\sun}$ models, CNO-ignition occurs while the hydrogen mass fraction at the centre is still equal to its initial value (0.75).}
\label{hr60}
\end{figure*}

Figure~\ref{hr60} shows the HR diagram for our 60 M$_{\sun}$ models. During the MS, the shift of the tracks in $T_{\rm eff}$ are slightly reduced compared to the 15 $M_{\sun}$ models: by 0.18 dex. Also, all the 60 $M_{\sun}$ models are directly sustained by the CNO cycle on the ZAMS, so the tracks are just shifted regularly, without affecting the shape of the tracks. During CHeB, however, the behavior we described for the 15 $M_{\sun}$ models with $\delta_{NN} < 0$ is more pronounced in the case of the 60 $M_{\sun}$ models: $^{12}$C and $^{16}$O are already exhausted at the end of CHeB (case III) for $\delta_{NN} \le -0.005$. This can be understood because the $^{12}$C($\alpha$,$\gamma$)$^{16}$O, the $^{16}$O($\alpha$,$\gamma$)$^{20}$Ne and the $^{20}$Ne($\alpha$,$\gamma$)$^{24}$Mg reaction rates, are a factor of 10 to 100 higher than the 3$\alpha$ rate when log $T_{\rm c} \approx 8.48$, \textsl{i.e.} when there is still about 5\% of helium in the core. Instead of a CO core, these models are left with an almost pure $^{24}$Mg core.

\begin{figure}
\centering
 \resizebox{.8\hsize}{!}{\includegraphics{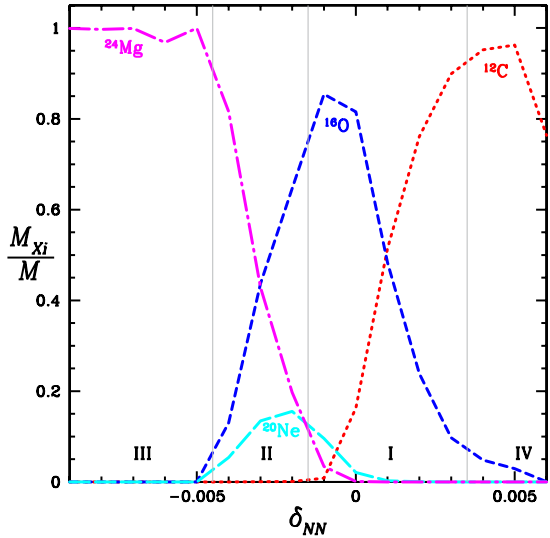}}
\caption{The composition of the core at the end of the central He burning in the 60 $M_{\sun}$ models as a function of $\delta_{NN}$.}
\label{sum60}
\end{figure}

For the 60 $M_{\sun}$ models with $\delta_{NN} > 0$, there is still a reasonable abundance of oxygen up to $\delta_{NN}$ = +0.003. At higher values of $\delta_{NN}$, we are again left with a nearly pure carbon core. The model with $\delta_{NN} = +0.006$ has proven to be very difficult to follow at the end of CHeB and was stopped before complete He exhaustion. The results for the 60 M$_{\sun}$ models are summarized in Fig.~\ref{sum60} which shows the composition of the core at the end of the CHeB phase. As in the  case of the 15 M$_{\sun}$ models, one can clearly see the strong dependence of the core composition on $\delta_{NN}$.

The effect of varying $\delta_{NN}$ on the core size is less clear in the case of the 60 $M_{\sun}$ models. In some cases, the model undergoes a CNO boost in the H-burning shell during CHeB, which reduces the core mass\footnote{We refer the interested reader to \citet{hirschi07} or \citet{emchm08} for a more detailed description of this phenomenon.}. The occurance of the boost does not follow a clear trend with $\delta_{NN}$. It appears on the HR diagram as a sudden drop in luminosity and effective temperature in the redwards evolution during CHeB (see Fig.~\ref{fdhrtc15}, \textsl{left}).

\subsection{Limits on the variation of the fundamental constants}

All of the models considered were followed without any numerical or evolutionary problem through the MS. The differences in lifetimes and tracks during this phase are not constraining enough to allow the exclusion of some range in $\delta_{NN}$ between -0.009 and +0.006. However, the CHeB phase amplifies these differences. 

At the end of CHeB, the models with $\delta_{NN} \le -0.005$ for the 15 M$_{\sun}$ model and $\delta_{NN} \le-0.004$ for the 60 M$_{\sun}$ model have virtually no $^{12}$C in the core, which means that the ``standard'' succession of stellar evolution burning phases will not be respected (see the bottom right panel of Fig.~\ref{fxco1}). These models are also devoid of $^{16}$O or $^{20}$Ne as well, leaving us with a nearly pure $^{24}$Mg core. Note that at this phase, the central temperature is close to that which would allow the $^{24}$Mg($\gamma$,$\alpha$)$^{20}$Ne or $^{24}$Mg($\alpha$,$\gamma$)$^{28}$Si reactions to take place. Therefore, there is a possibility that the nucleosynthetic chain could go on despite its strange evolution. However, the Geneva code is developed to follow the standard phases of stellar evolution, making it necessary to be modified before being able to follow further the evolution of these odd objects.

The models with $\delta_{NN}$ between -0.002 and -0.005 (between -0.002 and -0.004 for the 60 M$_{\sun}$ model) end the CHeB phase with a central abundance of $^{12}$C between 10$^{-4}$ and 10$^{-7}$, which means that the central C-burning phase will be extremely short. The $^{20}$Ne abundance at that stage is comprised between 0.04 and 0.10, so there will be a short phase of neon photodisintegration. Moreover, the $^{16}$O abundance ranges between 0.94 and 0.44 so the oxygen fusion phase will be almost normal. While the succession of the burning phases seems preserved,  one can however suppose that these models will present very different yields than the standard case with $\delta_{NN} = 0$. This point could be the subject of a future study. It is interesting to note here that since the C-burning phase is very short (due to the very low $^{12}$C abundance at the end of CHeB), the model will not have much time to lose entropy by neutrinos losses. We can suppose that the iron core will be hotter and bigger, so the remnant could be a black hole instead of a neutron star \citep{WW86,schlattl2004}.

The models with $\delta_{NN} > 0$ end the CHeB phase with larger and larger $^{12}$C abundances for increasing $\delta_{NN}$. The carbon burning phase will thus be much longer for these models which will lose a lot of energy through neutrino emission. A more suspicious feature is that the $^{16}$O production becomes negligible or even null for $\delta_{NN} \geq +0.003$ (see the top right panel of Fig.~\ref{fxco1}) ($\ge +0.004$ for the 60M$_{\sun}$ model). Normally the bulk of the $^{16}$O production occurs during CHeB: during C burning, the $^{16}$O abundance is reduced by $^{16}$O($\alpha$,$\gamma$)$^{20}$Ne, and during Ne burning, only a small fraction is produced by the photo-disintegration reaction $^{20}$Ne($\gamma$,$\alpha$)$^{16}$O. It would thus mean that such stars do not produce any $^{16}$O. This would pose difficulties for explaining the high O overabundances observed in extremely iron-poor stars found in the Galactic halo \citep[see][]{frebel08}.

From the preceding discussion, if we exclude a core composed exclusively of $^{24}$Mg (case III), we must reject $\delta_{NN} < -0.005$ for the 15 M$_{\sun}$ model. If we consider that a core only composed of $^{12}$C is not acceptable either (case IV), we must reject $\delta_{NN} > +0.002$. If we consider that a reasonable value of C/O must lay close to unity, we must also reject case II and the allowed range for $\delta_{NN} $ is further restricted to -0.001  to +0.002. Similarly for the 60 M$_{\sun}$ model, excluding cases III and IV leads to a limit $-0.004 \leq \delta_{NN} \leq +0.003$. The more stringent condition on C/O $\sim$ 1 leads to $-0.001 \leq \delta_{NN} \leq +0.003$.

\section{Discussion}

As we have seen in the previous sections, the extreme sensitivity of the triple $\alpha$ process to the resonant energy of the Hoyle state can lead to very different histories for massive Population III stars.  In particular, we have shown that very slight variations in the nucleon-nucleon potential (of order a few $\times 10^{-3}$) can lead to very different core compositions at the end of CHeB. We identified two cases (III and IV) corresponding to nearly pure $^{24}$Mg or pure $^{12}$C cores. These cases were present in both the 15 and 60 M$_{\sun}$ models studied. Below $\delta_{NN} = -0.005$, the stars end the CHeB phase with a core that is almost completely deprived of carbon, oxygen and neon. This is due to the $^{12}$C production by the 3$\alpha$ reaction becoming extremely weak compared to the $^{12}$C($\alpha$,$\gamma$)$^{16}$O reaction \citep[for which we have used the rates of][]{kunz02}. As soon as a little amount of $^{12}$C is produced, it is transformed into $^{16}$O, which in turn is transformed into $^{20}$Ne and then $^{24}$Mg due to the high temperature and density at which He burning occurs in these models. Above $\delta_{NN} = +0.002$, the models end the CHeB phase devoid of $^{16}$O.

We have checked the limiting values for $\delta_{NN}$ variation with stellar models in two different mass domains. For the 15 $M_{\sun}$ models, the lower limit is slightly larger than for the 60 $M_{\sun}$ models. This is due to the fusion phases occurring at higher $T_{\rm c}$ in the more massive stars, at conditions where the $^{12}$C($\alpha$,$\gamma$)$^{16}$O, $^{16}$O($\alpha$,$\gamma$)$^{20}$Ne and $^{20}$Ne($\alpha$,$\gamma$)$^{24}$Mg reaction rates are largely dominant over the 3$\alpha$ rate. A weak 3$\alpha$ reaction is a bigger handicap in the high mass domain. In contrast,  the upper limit is larger for the 60 $M_{\sun}$ models, because the 3$\alpha$ reaction rate is a little less extreme at higher $T_{\rm c}$.

Excluding these cases allows us to set a relatively conservative limit on  $\delta_{NN}$,
\begin{equation}
-0.004 < \delta_{NN} < +0.002 \, .
\label{weak}
\end{equation}
A more aggressive limit would also exclude case II in which CHeB ends with a  $^{16}$O and $^{20}$Ne core with little or no $^{12}$C. In this case, one could argue
\begin{equation}
-0.001 < \delta_{NN} < +0.002 \, .
\end{equation}
For the remainder of the discussion, we will restrict our attention to the weak limit (\ref{weak}), as our conclusions can be easily scaled to the stronger limit.

The limit in Eq. (\ref{weak}) stems directly from the variation in $Q_{\alpha\alpha\alpha}$.
Excluding regions III and IV amount to limiting $Q_{\alpha\alpha\alpha}$ to a range 
0.3142 -- 0.5100 MeV, or
\begin{equation}
-0.17 < \frac{\Delta Q_{\alpha\alpha\alpha}}{Q_{\alpha\alpha\alpha}} < 0.34.
\end{equation}

As discussed in \S \ref{sens}, a variation in $\delta_{NN}$ will result in a variation in the deuterium binding energy. Using Eq. (\ref{BDdNN}), the bound (\ref{weak}) thus becomes
\begin{equation}
-0.023 < {\Delta B_D \over B_D}  < +0.011 \, .
\label{bdbound}
\end{equation}
In principle, one would like to next convert the limit on $\delta_{NN}$  or $B_D$ into a limit on the fundamental constants. Unfortunately, as we have argued earlier, i) the direct limit from the triple $\alpha$ process based on $\delta_\alpha$ is far weaker than that due to $\delta_{NN}$ and ii) in the absence of some guiding theory of unification, we can not relate the variation of $B_D$ directly to a variation in $\alpha_{em}$. However, as discussed above and in more detail in \citet{coc07}, we can use gauge coupling unification to relate a variation in $B_D$ to a variation in $\alpha_{em}$ through $\Lambda$.  Ignoring first any variation in the Yukawa couplings and Higgs vev, thus using $\Delta B_D/ B_D = 18 R \Delta \alpha_{em} / \alpha_{em}$, with $R = 36$ as is expected in the simplest grand unified theories, we obtain
\begin{equation}
-3.5 \times 10^{-5} < {\Delta \alpha_{em} \over \alpha_{em}}  < +1.8 \times 10^{-5} \, .
\label{abound1}
\end{equation}
If we further assume the relations between gauge and Yukawa couplings and use Eq. (\ref{DeltaBd4}), the limit, though more speculative, is actually weakened by a factor of about 2 due to the partial cancellation between the gauge and Yukawa contributions to $B_D$. 

The limits on the variation of the fine structure constant derived above corresponds to a variation in $\alpha_{em}$ between the present time and a period around a redshift $z \sim 15-20$ where the Population III stars would have been present. These values are compatible with the similar limits (also assuming gauge coupling unification) on the variation of $\alpha_{em}$ at a redshift of 10$^{10}$ from BBN predictions.  They are larger by a factor of 10 than the values found in the claimed detections \citep{webb01,murph03,murphy07}  or a factor of 10 weaker than the limits from the non-detection \citep{chand,sri04,quast04,srianand2007} of a variation in $\alpha_{em}$ from quasar absorption systems at redshifts $z \leq 3.5$.

We remind the reader, that in the present work, the variation in $\delta_{NN}$ is only taken into account for the 3$\alpha$ reaction. If the other rates are also affected, the limits found here would potentially have to be revised, because they have been determined by anomalies in the evolution that are due to a competition between the efficiency of the various rates. However, being a resonant reaction, the 3$\alpha$ reaction is expected to be the most sensitive. Following \citet{ober2000}, the $^{16}$O($\alpha$,$\gamma$)$^{20}$Ne reaction is not expected to be sensitive to variations in $\alpha_{em}$, while the $^{12}$C($\alpha$,$\gamma$)$^{16}$O could be more affected by such variations because of subthresholds in the $^{16}$O nucleus. According to the same authors, this last reaction is expected to be strengthened by a weakening of the nucleon-nucleon interaction. In this case, the effects described for cases II and III would be more dramatic than the ones presented here, and so the limits might be tighter.

We conclude by asking: is it reasonable to exclude values of $\delta_{NN}$  using nucleosynthetic constraints from stellar models? The criteria that we applied assumes the possibility for a ``normal'' succession of burning phases (H $\rightarrow$ He $\rightarrow$ C $\rightarrow$ Ne $\rightarrow$ O $\rightarrow$ Si). Although we have not done so here,  a modified code including ``non standard'' fusion phases, would allow us to follow these models further. It is expected that the resulting yields would present large anomalies. Given the current state of abundance determination in extremely metal poor stars, it is highly improbable that the first stars would not produce fair amounts of $^{12}$C and $^{16}$O. The conservative case seem thus to offer a reasonable limit on the variations of the fundamental constants.

\appendix
\setcounter{equation}{0}
\renewcommand{\theequation}{A\arabic{equation}}
\section{Details on the Microscopic model}\label{appmicro}

Here, we provide some technical details about the microscopic calculation used to determine the $^8$Be and $^{12}$C binding energies. This calculation is based on the description of the nucleon-nucleon interaction by the Minnesota (MN) force \citep{TLT77}, adapted to low-mass systems. 

The nuclear part of the interaction potential $V_N$ between nucleons $i$ and $j$ is given by
\begin{eqnarray}
V_{Nij}(r)&=&\left[V_R(r)+\frac{1}{2}(1+P^{\sigma}_{ij})V_t(r)
+\frac{1}{2}(1-P^{\sigma}_{ij})V_s(r)\right]\nonumber \\
&&\times \left[\frac{1}{2}u+\frac{1}{2}(2-u)P^{r}_{ij}\right],
\label{eq_a1}
\end{eqnarray}
where $r=|\mathbf{r}_i-\mathbf{r}_j|$ and $P^{\sigma}_{ij}$ and $P^{r}_{ij}$ are the spin and space exchange operators, respectively. The radial potentials $V_R(r),V_s(r),V_t(r)$ are expressed as Gaussians and have been optimized to reproduce various properties of the nucleon-nucleon system, such as the deuteron binding energy at $\it{\delta_{NN}}$ = 0, or the low-energy phase shifts. They have been fit as \citep{TLT77}
\begin{eqnarray}
V_R(r)&=&200\,\exp(-1.487r^2) \nonumber \\
V_s(r)&=&-91.85\,\exp(-0.465r^2) \nonumber \\
V_t(r)&=&-178\,\exp(-0.639r^2)
\label{vs}
\end{eqnarray}
where energies are expressed in MeV and lengths in fm.

In Eq.~(\ref{eq_a1}), the exchange-admixture parameter $u$ takes standard value $u=1$, but can be slightly modified to reproduce important properties of the $A$-nucleon system (for example, the energy of a resonance). This does not affect the physical properties of the interaction. The MN force is an effective interaction, adapted to cluster models. It is not aimed at perfectly reproducing all nucleon-nucleon properties, as realistic forces used in {\sl ab initio} models \citep{NQS09}, where the cluster approximation is not employed. The potentials are expressed as Gaussian factors, well adapted to cluster models, where the nucleon orbitals are also Gaussians \citep{WT77}.

The wave functions (\ref{eq4}) are written in the Resonating Group Method (RGM) which clearly shows the factorization of the system wave function in terms of individual cluster wave functions. In practice the radial wave functions are expanded over Gaussians, which provides the Generator Coordinate Method (GCM), fully equivalent to the RGM \citep{WT77} but better adapted to numerical calculations. Some details are given here for the simpler two cluster case. The radial function $g^{JM\pi}_2(\rho)$ is written as a sum over Gaussian functions centered at different values of the Generator Coordinate $R_n$. This allows us to write the $^8$Be wave function (\ref{eq4}) as
\begin{eqnarray}
\Psi^{JM\pi}_{^8Be}=\sum_n f^{J\pi}(R_n)\Phi^{JM\pi}(R_n),
\end{eqnarray}
where $\Phi^{JM\pi}(R_n)$ is a projected Slater determinant.

This development corresponds to a standard expansion on a variational basis. The binding energies $E^{J\pi}$ of the system are obtained by diagonalization of
\begin{eqnarray}
\sum_n \left[ H^{J\pi}(R_n,R_{n'})-E^{J\pi}N^{J\pi}(R_n,R_{n'})\right] f^{J\pi}(R_n)=0,
\end{eqnarray}
where the overlap and hamiltonian kernels are defined as
\begin{eqnarray}
N^{J\pi}(R_n,R_{n'})&=&\langle \Phi^{J\pi}(R_n) | \Phi^{J\pi}(R_{n'}) \rangle, \nonumber \\
H^{J\pi}(R_n,R_{n'})&=&\langle \Phi^{J\pi}(R_n) |H| \Phi^{J\pi}(R_{n'}) \rangle .
\end{eqnarray}
The Hamiltonian $H$ is given by Eq.~(\ref{eq1}). Standard techniques exist for the evaluation of these many-body matrix elements \citep{Br66}. The choice of the nucleon-nucleon interaction directly affects the calculation of the hamiltonian kernel, and therefore of the eigenenergy $E^{J\pi}$.

For three-body wave function, the theoretical developments are identical, but the presentation is more complicated due to the presence of two relative coordinates $(\rho,R)$. The problem is addressed by using the hyperspherical formalism \citep{KD04}.

\setcounter{equation}{0}
\renewcommand{\theequation}{B\arabic{equation}}
\section{Reaction rates  and numerical integration}
\label{s:rate}

To take into account the (energy dependent) finite widths of the two resonances involved in this two step process, one has to perform numerical integrations as was done in NACRE following \citet{Nom85} and \citet{Lan86}. Here, the condition of thermal equilibrium is relaxed, but it is assumed that the time scale for alpha capture on $^8$Be is negligible compared to its lifetime  against alpha decay.
The rate is calculated as in NACRE for the resonance of interest:
\begin{eqnarray*}
N_{\rm A}^2 \langle \sigma v \rangle^{\alpha\alpha\alpha} & = & 
3N_{\rm A}
\left( \frac{8\pi\hbar}{\mu_{\alpha\alpha}^2} \right)
\left( \frac{\mu_{\alpha\alpha}}{2\pi k_{\rm B} T} \right)^{3/2}\times\\
\end{eqnarray*}
\begin{eqnarray}
\int_0^{\infty}
\frac{\sigma_{\alpha\alpha}(E)}{\Gamma_{\alpha}(E)}
\exp(-E/k_{\rm B} T) N_{\rm A} \langle \sigma v \rangle^{\alpha ^8{\rm Be}}
\, E \, dE,
\end{eqnarray}
where $\mu_{\alpha\alpha}$ is the reduced mass of the $\alpha$ + $\alpha$ system, and $E$ is the energy with respect to the $\alpha$ + $\alpha$ threshold. The elastic cross section of $\alpha$ + $\alpha$ scattering is given by a Breit-Wigner expression:
\begin{equation}
\sigma_{\alpha\alpha}(E)\, = \,
\frac{\pi}{k^2} \, \omega \, 
\frac{\Gamma_\alpha^2(E)}
{(E-E_R))^2+\Gamma_\alpha^2(E)/4},
\label{eq:bw}
\end{equation}
where  $k$ is the wave number, 
$E_R{\equiv}E_R(^8{\mathrm Be})$, $\Gamma_\alpha{\equiv}\Gamma_\alpha(^8{\mathrm Be})$, $\omega$ is a statistical factor (here equal to 2 to account for identical particles with spin zero).

The $N_{\rm A} \langle \sigma v \rangle^{\alpha^8{\rm Be}}$ rate assumes that $^8$Be has been formed at an energy $E$ different from $E_{^8 \rm Be}$ \citep{Lan86}. This rate is given by
\begin{eqnarray*}
N_{\rm A} \langle \sigma v \rangle^{\alpha ^8{\rm Be}} =
N_{\rm A} \frac{8\pi}{\mu_{\alpha ^8{\rm Be}}^2}
\left( \frac{\mu_{\alpha ^8{\rm Be}}}{2\pi k_{\rm B} T} \right)^{3/2}\times
\end{eqnarray*}
\begin{eqnarray}
\int_0^{\infty} \sigma_{\alpha ^8{\rm Be}}(E';E)
\exp(-E'/k_{\rm B} T)\ E'\, dE',
\end{eqnarray}
where $\mu_{\alpha ^8{\rm Be}}$ is the reduced mass of the $\alpha$ + $^8$Be system, and $E'$ is the energy with respect to its threshold (which varies with the formation energy $E$). As in \citet{Nom85,Lan86}, we parametrize $\sigma_{\alpha ^8{\rm Be}}(E';E)$ as
\begin{eqnarray*}
\sigma_{\alpha ^8{\rm Be}}(E';E)= 
\frac{\pi\hbar^2}{2\mu_{\alpha ^8{\rm Be}}E'} \times
\end{eqnarray*}
\begin{eqnarray}
\frac{\Gamma_{\alpha}(E')
\Gamma_{\gamma}(E'+E)}
{[E' - E_R(^{12}{\mathrm C}) + E - E_R(^8{\mathrm Be})]^2 + \frac{1}{4}
\Gamma (E';E)^2},
\end{eqnarray}
where the partial widths are those of the Hoyle state and in particular, $\Gamma=
\Gamma_\alpha({}^{12}{\mathrm C})+\Gamma_\gamma(^{12}{\mathrm C})$. The various integrals are calculated numerically. The experimental widths at resonance energy can be found in Table~\ref{t:widths}. 

However, one must include $i$) the energy dependence of those widths, away from the resonance energy and $ii$) the variation of the widths at  the resonant energy when this energy changes due to a change in the nuclear interaction.  

The energy dependence of the particle widths $\Gamma_\alpha(E)$ is given by:
\begin{equation}
\Gamma_\alpha (E) 
= \Gamma_\alpha (E_R) \ \frac{P_{\ell}(E,R_c)}{P_{\ell}(E_R,R_c)},
\end{equation}
where $P_{\ell}$ is the penetration factor associated with the relative angular momentum $\ell$ (0 here) and the channel radius, $R_c$\footnote{We choose $R_c$ = $1.3 \, (A_1^{1/3} + A_2^{1/3})$ fm, for nuclei $A_1$ and $A_2$.}. The penetration factor is related to the Coulomb functions by:
\begin{equation}
P_{\ell}(E,R) = {{\rho}\over{F_\ell^2(\eta,\rho)+G_\ell^2(\eta,\rho)}}
\label{eq:penet}
\end{equation}
where $\rho=kR$ and 
\begin{equation}
\eta={{Z_1Z_2\alpha_{em}}\over{v/c}}
\label{eq:eta}
\end{equation}
is the Sommerfeld parameter.

For radiative capture reactions, the energy dependence of the gamma width
$\Gamma_{\gamma}(E)$ is given by:
\begin{equation}
 \Gamma_{\gamma}(E){\propto}\alpha_{em}E^{2\lambda+1}
\end{equation}
where $\lambda$ is the multipolarity (here 2 for $E2$) of the electromagnetic transition. 

\begin{figure}
\centering
 \resizebox{.8\hsize}{!}{\includegraphics{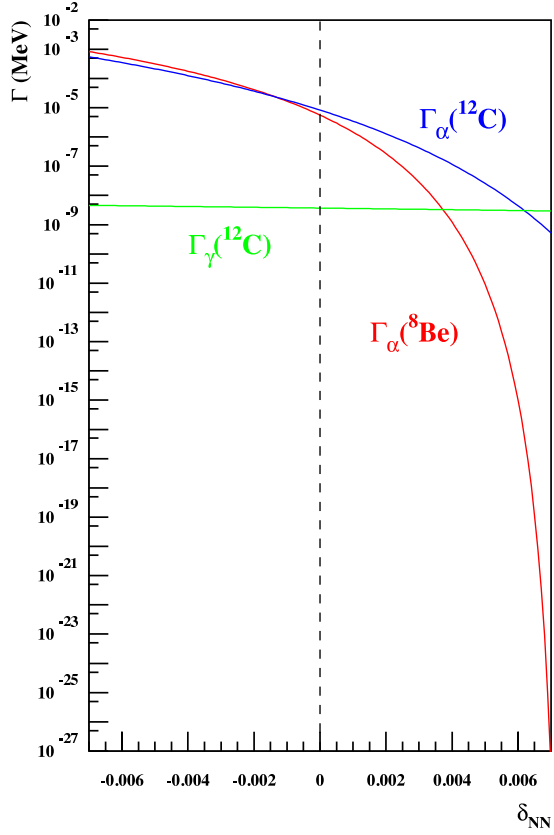}}
\caption{The  partial widths of ${}^8$Be and ${}^{12}$C as a function of $\delta_{NN}$.}
\label{fwidths}
\end{figure}

The relevant widths as a function of $\delta_{NN}$ are given in Figure~\ref{fwidths}. They are directly linked to the resulting change of $E_R$($^8$Be) and $E_R$($^{12}$C).

The radiative width, $\Gamma_\gamma(^{12}{\mathrm C})$ with its $E^5$ energy dependence shows little evolution. (The energy of the final state at 4.44~MeV is assumed to be constant). In contrast, the $^8$Be alpha width undergoes large variations due to the effect of Coulomb barrier penetrability. Note that compared to these variations, those induced by a change of $\alpha_{em}$ in the Coulomb barrier penetrability (Eqs.~\ref{eq:eta}, \ref{eq:penet}) and $\Gamma_\gamma$ are considerable smaller.

Numerical integration is necessary at low temperature as the reaction takes place through the low energy wing of resonances. It takes even more relative importance, at a given temperature, when the resonance energy is shifted upwards. On the other hand, when $\delta_{NN}$ increases, the resonance energies decrease, and the $\Gamma_\alpha(^8{\mathrm Be})$ becomes so small that the numerical integration becomes useless and soon gives erroneous results because of the finite numerical resolution. For this reason, when $\Gamma_\alpha(^8{\mathrm Be})\ <\ 10^{-8}$~MeV, we use instead the Saha equation for the first step and the sharp resonance approximation for the second step, i.e. Eq.~(\ref{eq:anal}) when $\Gamma_\alpha(^{12}{\mathrm C})\ <\ 10^{-8}$~MeV. (Note that for high values of  $\delta_{NN}$, the condition $\Gamma_\gamma\ll\Gamma_\alpha$ does not hold anymore and $\gamma\neq\Gamma_\gamma$.)

At temperatures in excess of $T_9 \simeq 2$, one must include the contribution of the higher $^{12}$C levels like the one
observed by \cite{Fynbo}. As this is not of importance for this study, we just added the contribution given by the last terms in the NACRE analytical approximation and neglected any induced variation. 

\begin{figure}
\centering
 \resizebox{\hsize}{!}{\includegraphics{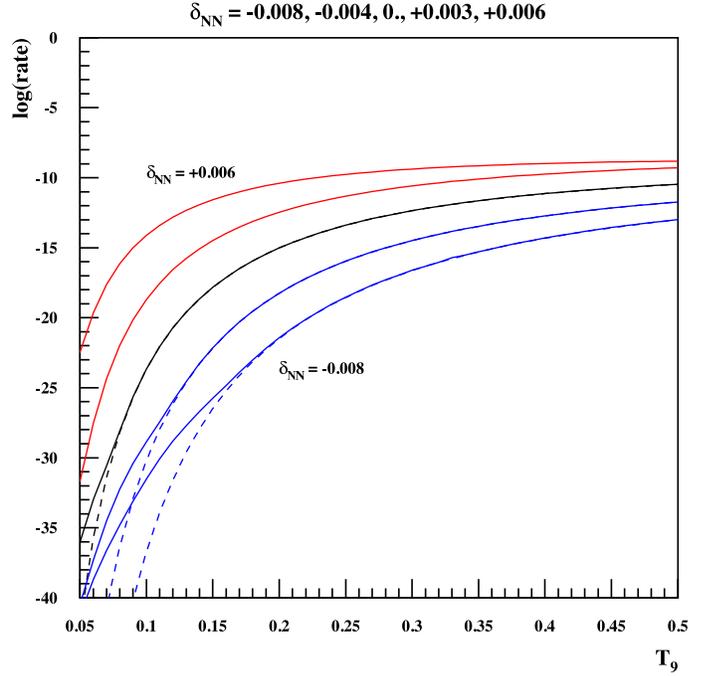}}
\caption{The $^4$He($\alpha\alpha,\gamma$)$^{12}$C reaction rate as a function of temperature for different values of $\delta_{NN}$. Solid (dashed) lines represent the result of the numerical calculation (analytical approximation) with $\delta_{NN}$=0 (black), $\delta_{NN}\ >$ 0 (red) and $\delta_{NN}\ <\ $0 (blue). (For the negative value of $\delta_{NN}$, the larger difference is caused by the failure of the numerical integration and the analytical solution is preferred (see text).)
}
\label{frates}
\end{figure}

Figure~\ref{frates} shows the numerically integrated $^4$He($\alpha\alpha,\gamma$)$^{12}$C reaction rates for different values of $\delta_{NN}$ compared with the analytical approximation (Eq.~(\ref{eq:anal})). The difference is important at low temperature and small $\delta_{NN}$ values but becomes negligible for $\delta_{NN}$ $\gtrsim$ 0. At the highest values of $\delta_{NN}$ we consider, the numerical calculation uses the Saha equation for the first step but the total widths of the $^{12}$C level becomes also too small to be accurately numerically calculated: we use Eq.~(\ref{eq:anal}) instead.   

The $^8$Be lifetime w.r.t. alpha decay, ($h/\Gamma_\alpha(^8{\mathrm Be})$), exhibits the opposite behavior indicating that for large values of $\delta_{NN}$ it becomes stable. Before that, its lifetime is so long that the $^4$He($\alpha\alpha,\gamma$)$^{12}$C reaction should be considered as a real two step process with $^8$Be included in the network as the assumption that alpha decay is much faster than alpha capture may not hold anymore. Fortunately, our network calculations shows that this situation is encountered only for  $\delta_{NN}\gtrsim$ 0.006 for the temperatures and densities considered in our stellar evolution studies.

\acknowledgement{This work was partly supported by i) CNRS PEPS/PTI, ii) CNRS PICS France--USA
and by DOE grant
DE-FG02-94ER-40823 at the University of Minnesota.}

\bibliographystyle{aa}
\bibliography{finestruc}

\end{document}